\documentclass[twocolumn,pra
,superscriptaddress
]{revtex4}

\usepackage{graphicx}
\usepackage{amssymb}
\usepackage{amsmath}
\usepackage{color}
\usepackage{comment}

\newcommand{\vect}[1]{{\mathbf #1}}
\newcommand{\up}{\uparrow}
\newcommand{\down}{\downarrow}
\renewcommand{\k}{{\bf k}}
\newcommand{\p}{{\bf p}}
\newcommand{\q}{{\bf q}}

\newcommand{\0}{{\bf 0}}

\newcommand{\ad}{a_{2\rm{D}}}
\newcommand{\bra}[1]{\langle\left.{#1}\right|}
\newcommand{\ket}[1]{\left|{#1}\right.\rangle}
\newcommand{\ef}{E_{F\uparrow}}
\newcommand{\kf}{k_{F\uparrow}}
\newcommand{\eb}{\varepsilon_B}
\newcommand{\eq}{\xi_{\q\up}}
\newcommand{\ek}{\xi_{\k\up}}

\newcommand{\ekone}{\xi_{\k_1\up}}
\newcommand{\ektwo}{\xi_{\k_2\up}}
\newcommand{\eqone}{\xi_{\q_1\up}}
\newcommand{\eqtwo}{\xi_{\q_2\up}}
\newcommand{\ekN}{\xi_{\k_N\up}}
\newcommand{\eki}{\xi_{\k_i\up}}
\newcommand{\nn}{\nonumber}

\newcommand{\fs}{\ket{FS}}

\def\lba{\left(}    \def\rba{\right)}
\def\lbc{\left[}    \def\rbc{\right]}

\def \del{\partial}    % for writing partial derivatives

\begin{document}

%\linenumbers

%\title{Many-body bound states in two-dimensional atomic gases}
\title{Highly polarized Fermi gases in two dimensions}
%\title{Single ``impurity'' atom in a two-dimensional atomic Fermi gas}
%\title{Single impurity atom in a two-dimensional atomic Fermi gas}

\author{Meera M. Parish}
%\email{meera.parish@ucl.ac.uk} %
\affiliation{London Centre for Nanotechnology, Gordon Street, London, WC1H 0AH, United Kingdom}
\affiliation{%T.C.M. Group, 
Cavendish Laboratory, JJ Thomson Avenue, Cambridge,
 CB3 0HE, United Kingdom} %

\author{Jesper Levinsen}
%\email{} %
\affiliation{%T.C.M. Group, 
Cavendish Laboratory, JJ Thomson Avenue, Cambridge,
 CB3 0HE, United Kingdom} %

\date{\today}

\begin{abstract}
  We investigate the highly polarized limit of a two-dimensional (2D)
  Fermi gas, where we effectively have a single spin-down impurity
  atom immersed in a spin-up Fermi sea.  By constructing variational
  wave functions for the impurity, we map out the ground state phase
  diagram as a function of mass ratio $m_\uparrow/m_\downarrow$ and
  interaction strength. In particular, we determine when it is
  favorable for the dressed impurity (polaron) to bind particles from
  the Fermi sea to form a dimer, trimer or even larger clusters.
  Similarly to 3D, we find that the Fermi sea favors the trimer state
  so that it exists for $m_\uparrow/m_\downarrow $ less than the
  critical mass ratio for trimer formation in the vacuum.  We also
  find a region where dimers have finite momentum in the ground state,
  a scenario which corresponds to the Fulde-Ferrell-Larkin-Ovchinnikov
  superfluid state in the limit of large spin imbalance.  For equal
  masses ($m_\up = m_\down$), we compute rigorous bounds on the
  polaron-dimer transition, and we show that the polaron energy and
  residue is well captured by the variational approach, with the
  former quantity being in good agreement with experiment.  When there
  is a finite density of impurities, we find that this polaron-dimer
  transition is preempted by a first-order superfluid-normal
  transition at zero temperature, but it remains an open question what
  happens at finite temperature.
\end{abstract}

\pacs{03.75.Ss, 67.85.-d, 64.70.Tg}

\maketitle

\section{Introduction}
The creation of tunable Fermi systems with ultracold atomic gases has
greatly renewed interest in fermionic pairing phenomena and
superfluidity. In particular, recent experiments have successfully
confined fermionic atoms to a quasi-two-dimensional (quasi-2D)
geometry~\cite{2DFermi_expt,PhysRevLett.106.105301,Dyke2011,2011Natur.480...75F,
  sommer2011_2D,Koschorreck2012,Zhang:2012uq}, thus enabling the
investigation of 2D Fermi gases.  Such model 2D systems can
potentially provide insight into more complicated solid state systems
such as electron-hole bilayers~\cite{Croxall2008,Seamons2009} and high
temperature superconductors~\cite{Norman2011}.  Moreover, 2D is the
``marginal'' dimension where quantum fluctuations are enhanced
compared to 3D, leading to the destruction of Bose Einstein
condensation (BEC) at finite temperature. At the same time, there is a
dearth of exactly solvable models, in contrast to the case in 1D.
This makes the cold-atom system even more important for testing
theoretical approaches in 2D.

A canonical many-body problem currently under investigation is that of
an impurity interacting with a medium. This so-called ``polaron''
problem traditionally involves a bosonic medium such as a bath of
phonon excitations in a crystal~\cite{frohlich1954}.  However, the
advent of cold atoms has extended this problem to include a fermionic
medium, where the impurity can now undergo a sharp transition and
effectively change its statistics by binding fermions from the
background Fermi gas~\cite{prokofiev2008,prokofiev2008_2}. For a
fermionic impurity, this corresponds to the highly polarized limit of
a two-component ($\up$, $\down$) Fermi gas and, thus, the existence of
such binding transitions will determine the topology of the phase
diagram for the spin-imbalanced Fermi
gas~\cite{parish2007,sheehy2007}.  In 3D, the $\down$ impurity is
initially dressed by density fluctuations of the $\up$ Fermi gas,
forming a polaron state~\cite{chevy2006_2}, but with increasing
attractive interactions it can bind an extra $\up$ particle to form a
dimer
(molecule)~\cite{prokofiev2008,prokofiev2008_2,combescot2009,mora2009,punk2009,Mathy:2011ys}
or it can even bind two $\up$ particles to form a
trimer~\cite{Mathy:2011ys}, depending on the mass ratio
$m_\up/m_\down$. For the 2D case, there has been debate about whether
strong quantum fluctuctions preclude the existence of such binding
transitions~\cite{Zollner:2011fk,Klawunn:2011fk}, but one of us has
recently argued that sharp polaron-molecule transitions can
exist~\cite{Parish:2011vn} and this appears consistent with
experiment~\cite{Koschorreck2012,Levinsen2012}.

In this work, we further investigate this issue by mapping out the
ground state phase diagram for the 2D impurity problem as a function
of mass ratio $m_\up/m_\down$ and interaction strength.  Similarly to
3D, we show that the impurity can bind one, two or more $\up$
particles from the Fermi sea to form bound clusters dressed by density
fluctuations. We approximate the different impurity states using
variational wave functions that include a finite number of
particle-hole excitations of the Fermi sea. We find that the presence
of a Fermi sea appears conducive to the formation of trimers, which in
vacuum exist for $m_\up/m_\down>3.33$~\cite{2Dtrimer}. On the other
hand, tetramers containing three $\up$ particles and existing for
$m_\up/m_\down>5.0$ in vacuum~\cite{Levinsen2013}, appear to be
disfavored in the many-body system. We further find a region of
parameter space in which the impurity binds a single particle from the
Fermi sea to form a dimer with finite momentum.  A finite density of
such molecules will form a spatially modulated condensate and thus
this state is the single-particle analog of the elusive
Fulde-Ferrell-Larkin-Ovchinnikov (FFLO)
phase~\cite{PhysRev.135.A550,1965JETP...20.762L}.

The presence of bound clusters of $N+1$ particles such as trimers and
tetramers ($N=2$ and 3) introduce important $(N+1)$-body
correlations. Even if these states are metastable, their presence can
still affect the system: for instance, we expect three-body
correlations to be more important in 2D than in 3D for the
mass-balanced system due to the smaller repulsive barrier between the
identical fermions, and the resulting enhanced three-body
interactions~\cite{PhysRevLett.103.153202,Ngampruetikorn2013}. In
order to quantify the effect of these higher order correlations, we
consider the polaron state dressed by two particle-hole pairs for
equal masses ($m_\up=m_\down$). This wavefunction includes three-body
correlations, whereas the usual Chevy wavefunction~\cite{chevy2006_2}
contains only two-body correlations.  The results of these two
variational wavefunctions match in the limit of weak interactions and
we also find a reasonable agreement across the regime of strong
interactions. This approach further allows us to arrive at rigorous
upper and lower bounds for the position of the polaron-molecule
transition, yielding $-0.97<\ln(\kf\ad)<-0.802$, with $\kf$ the Fermi
momentum of majority $\up$ particles and $\ad$ the 2D scattering
length.

At any given interaction strength, only one state may be the ground
state and all other states are at best metastable. The polaron
wavefunction of Ref.~\cite{chevy2006_2} leads to a completely real
equation for the polaron's energy, and thus one might think that the
variational approach may only be used to calculate ground state
properties of the polarized Fermi gas. On the other hand, using a
diagrammatic approach as in Ref.~\cite{Combescot:2007bh} to describe
the polaron, naturally incorporates an imaginary part in the energy,
and thus the metastability of excited states is automatically
present. We resolve this seeming discrepancy by showing here that a
straightforward change of the variational wavefunction and the
minimization procedure allows the variational approach to be extended
to describe not only the ground state but also excited branches. Thus
the two approaches are, in fact, fully equivalent.

The experimental observation of the transition from a polaron to a
molecule may be precluded by phase separation into an unpolarized
superfluid phase and a fully polarized normal phase. Using results
from a quantum Monte Carlo (QMC) simulation of the superfluid phase in
an equal-mass unpolarized Fermi gas~\cite{PhysRevLett.106.110403}, we
indeed find this to be the case in the highly-polarized
limit. However, this result only strictly applies to zero temperature
and it remains an open question whether this occurs at finite
temperature.

The paper is organized as follows: The contact interaction is
introduced for the 2D system in
Sec.~\ref{sec:prelim}. Sections~\ref{sec:bare} and \ref{sec:dressed}
introduce the variational wave functions used to describe the possible
ground states of the impurity atom, while Sec.~\ref{sec:phase}
describes the resulting single-impurity phase diagrams. In
Sec.~\ref{sec:dens} we derive the condition for phase separation
precluding the polaron-molecule transition, while in
Sec.~\ref{sec:2ph} we introduce the polaron dressed by two
particle-hole pairs. In Sec.~\ref{sec:meta} we show the manner in
which the variational method may be modified to describe metastable
states, and finally in Sec.~\ref{sec:conclusion} we conclude.

\section{Preliminaries \label{sec:prelim}}

We consider the Hamiltonian for a two-component atomic Fermi gas
interacting via a short-range interaction in two dimensions (2D):
\begin{align}\label{eq:Ham}
 H = & \sum_{\vect{k}\sigma} \epsilon_{\vect{k}\sigma}
 c^\dag_{\vect{k}\sigma}c_{\vect{k}\sigma}
 +g% \frac{g}{\Omega}
\sum_{\vect{k},\vect{k'},\vect{q}}
 c^\dag_{\vect{k}\uparrow}c^\dag_{\vect{k'}\downarrow}
 c_{\vect{k'}+\vect{q}\downarrow}c_{\vect{k}-\vect{q}\uparrow}~,
\end{align}
where the spin $\sigma=\up,\down$, $\epsilon_{\vect{k}\sigma} =
\frac{\vect{k}^2}{2m_\sigma}$, and $g$ is the strength of an
attractive contact interaction. We work in units where $\hbar$ and the
system area are both 1.  Note that since we are considering
low-energy, $s$-wave interactions, the Pauli exclusion suppresses
interactions between the same species of fermion.  For the two-body
problem ($\uparrow$ and $\downarrow$), we simply have
\begin{align} \label{eq:2body}
-\frac{1}{g} & = %\frac{1}{\Omega}
\sum_{\vect{k}}^{\Lambda} 
\frac{1}{\eb + \epsilon_{\vect{k}\uparrow} + \epsilon_{\vect{k}\downarrow}},
\end{align}
where $\Lambda$ is the UV cut-off and $\eb$ is the binding energy of
the weakly bound diatomic molecule which always exists for an
attractive interaction in 2D. We see that the integral logarithmically
diverges if we fix $\eb$ and take $\Lambda \to \infty$. Thus $\Lambda$
cannot be removed from the problem and the binding energy $\eb$
depends on both $\Lambda$ and $g$, in contrast to 1D. For the
many-body system, our results become independent of the cut-off
$\Lambda$ once Eq.~\eqref{eq:2body} is used to replace $g$ with $\eb$.

At low energies, the elastic scattering of a spin-$\up$ and  a
spin-$\down$ atom at relative momentum $q$ is described by the
$s$-wave scattering amplitude \cite{LL}
\begin{equation}
f(q)=\frac{2\pi}{\ln\left[1/(q\ad)\right]+i\pi/2}.
\end{equation} 
Here, the momentum dependence of the scattering amplitude is
characterized by the 2D scattering length $\ad>0$. Analytic
continuation of the scattering amplitude to momenta on the positive
imaginary axis yields a pole in the scattering amplitude at
$q=i\ad^{-1}$. This corresponds to the binding energy
$\eb=1/2m_r\ad^2$, with the reduced mass defined as $m_r=m_\up
m_\down/(m_\up+m_\down)$.  Thus, the 2D scattering length essentially
corresponds to the size of the two-body bound state. Unlike in 3D, the
scattering amplitude in 2D does not reduce to a constant at low
scattering momenta; instead the scattering is strong in the regime
$\ln(q\ad)\sim0$ and weak when $|\ln(q\ad)|\gg1$. In the presence of a
Fermi sea, the characteristic momentum scale is set by the Fermi
momentum $k_{F}$ and we thus expect strong many-body effects when
$\ln(k_{F}\ad)\sim0$. Indeed, Bloom~\cite{PhysRevB.12.125} has
demonstrated how the 2D Fermi gas in the regime $|\ln(k_{F}\ad)|\gg1$
is perturbative in $1/\ln(k_{F}\ad)$.

In the following, we focus on the case where we have a single
spin-down minority atom immersed in a non-interacting Fermi sea of
spin-up atoms --- the extreme limit of population imbalance.  Note
that the phase diagram we obtain for the single impurity atom is
independent of impurity statistics and is thus also relevant to 2D
Bose-Fermi mixtures. However, the focus of this paper will be on
fermionic impurities. Defining the interaction parameter in the
imbalanced gas, $\eta\equiv\ln(\kf\ad)$, in the limit of weak
attractive interactions (or, equivalently, the large-density limit)
where $\eta \gg 1$, the ground state is expected to approach that of
the non-interacting system: $\ket{P_1(\0)} =
c^\dag_{\0\downarrow}\fs$, where $\fs$ represents the Fermi sea of
$\uparrow$-particles. Note that the subscript on the ``polaron'' wave
function $\ket{P}$ denotes the number of operators acting on the Fermi
sea; this is the nomenclature we will use throughout the paper.
Decreasing $\eta$ will eventually give rise to one or more binding
transitions, where the spin-down minority particle binds one or more
spin-up particles. The behaviour of the system in this regime can be
analysed with the use of variational wave functions for the different
bound states.

\section{``Undressed'' wave functions \label{sec:bare}}
In this section, we neglect the particle-hole excitations generated by
the impurity $\downarrow$-particle interacting with the Fermi sea and
focus on the ``bare'' part of the exact wave function.  For the
molecule and polaron, this is equivalent to the mean field approach
for the spin-imbalanced Fermi gas (see, e.g.,
Ref.~\cite{parish2007_2}).  In the present problem of a single
impurity interacting with a Fermi gas, the approach serves as a useful
introduction to the more complicated wavefunctions dressed by
particle-hole fluctuations.  It should also provide insight into the
relevant few-body correlations present in the system.  Below, in
Sec.~\ref{sec:dressed}, we consider wave functions dressed by one
particle-hole excitation, an approach which yields quantitatively
accurate results in the perturbative regimes of weak and strong
attractive interactions and which is also expected to give a good
description across the regime of strong 2D interactions $\eta\sim0$,
as we discuss below.

The results of this section become exact in the vacuum limit,
$\kf\to0$. In this case, we can study transitions between different
few-body states containing a single spin-$\down$ atom and $N$
spin-$\up$ particles. In vacuum, a bound diatomic molecule always
exists for two atoms interacting via contact interactions under a
strong two-dimensional confinement. Additionally, a trimer state
consisting of two heavy fermions and one light particle becomes
energetically favorable for a mass ratio $r\equiv m_\up/m_\down$ above
3.33~\cite{2Dtrimer}, while a tetramer containing three heavy fermions
and one light particle is favorable for $r>5.0$~\cite{Levinsen2013}.

\subsection{Molecules ($M_2$)}
The simplest, lowest-order variational wave function for a bound pair
or ``molecule'' is:
\begin{align}\label{eq:M2}
 \ket{M_2(\vect{p})} & = \sum_{\vect{k}} \varphi_{\vect{k}}^{(\vect{p})}  
c^\dag_{\vect{p}-\vect{k}\downarrow} c^\dag_{\vect{k}\uparrow} \fs,
\end{align}
where $\vect{p}$ corresponds to the center-of mass momentum of the
pair, while the spin-up particle momentum satisfies $k \equiv
|\vect{k}| > \kf$.  This wave function gives the exact two-body state
in the limit $\kf \to 0$ and is in fact identical to the BCS mean
field wave function for extreme imbalance (after the state with one
$\downarrow$ particle has been projected out). Note that we assume the
wave function \eqref{eq:M2} has one less $\up$ particle in the Fermi
sea compared with the non-interacting state $\ket{P_1}$ in order to
preserve particle number.  Indeed, we will assume throughout this
paper that all the impurity wave functions we introduce have the same
number of $\up$ particles as $\ket{P_1}$.  This is equivalent to
measuring the energy of $\up$ particles with respect to their chemical
potential, the Fermi energy $\ef \equiv \kf^2/2m_\up$, and thus we
define $\xi_{\k\up}\equiv\epsilon_{\k\up}-\ef$.

Minimizing the expectation value $\bra{M_2(\vect{p})} (H - E)
\ket{M_2(\vect{p})}$ with respect to $\varphi_{\vect{k}}^{(\vect{p})}$
yields an implicit equation for the molecule energy $E$:
\begin{align}
  -\frac1g&= %\frac{1}{\Omega}
\sum_{\vect{k}}%^{\Lambda}
  \frac{1}{-E% -\ef + \epsilon_{\vect{k}\uparrow} 
+\ek+
    \epsilon_{\vect{p}-\vect{k}\downarrow}}.
\end{align}
Here and in what follows we use the convention that the momentum $\k$
corresponds to a particle excited out of the Fermi sea, {\em i.e.}
$|\k|>\kf$. Additionally, the energy $E$ is assumed to be with respect
to the (macroscopic) energy of the non-interacting $\up$ Fermi gas.
We neglect Hartree terms involving $g \kf^2/4\pi$ since these vanish
when we take the limit $\Lambda \to \infty$, $g \to 0$. Converting
sums into integrals and sending $\Lambda \to \infty$ then gives:
\begin{widetext}
\begin{align}\notag
  2 \eb & = -E^\prime-\frac{p^2 (r-1)
    r}{2m_r(1+r)^2}+\frac{1}{1+r}\sqrt{\left(E^\prime
      +\left(E^\prime + \frac{\kf^2}{2m_r}-\frac{(p-\kf)^2}{2m_r}\right) r\right)
    \left(E^\prime+\left(E^\prime + \frac{\kf^2}{2m_r}-\frac{(p+\kf)^2}{2m_r}\right)
      r\right)}.
\end{align}
\end{widetext}
where we defined the energy $E^\prime = E - \frac{\kf^2}{2m_\down}$.
Clearly, when we approach the two-body limit ($\kf \to 0$), the
molecule has its lowest energy at zero momentum. For this case, we
simply have $E = \frac{\kf^2}{2m_\down} -\eb$.  However, once $\kf\ad >
1/\sqrt{r}$, we find that the molecule acquires a finite momentum:
\begin{align}\label{eq:M2_momentum}
p & = \frac{1+r}{r \ad} \sqrt{\kf \ad\sqrt{r}-1}.
\end{align} 
Increasing $\kf\ad$ further eventually causes the molecule to unbind
into the state $\ket{P_1(\0)}$.  At this unbinding transition
(assuming there is a direct transition), we always find that $p =
\kf$, in contrast to the 3D case where $p=0$ for mass ratios
sufficiently close to one~\cite{Mathy:2011ys}.  Referring to
Eq.~\eqref{eq:M2_momentum}, this means that the molecule unbinds when
$\kf \ad = (1+r)/\sqrt{r}$, i.e.\ when $\eb$ equals the center-of-mass
kinetic energy of the molecule at $p=\kf$.

\subsection{Trimers ($T_3$)}
Another possible bound state is the trimer consisting of two spin-up
fermions and one spin-down particle. Since this trimer involves
identical fermions, its angular momentum $L$ must necessarily be odd
and so the lowest-energy trimer is a $p$-wave ($L=1$) bound
state. Indeed, it may be regarded as a $p$-wave pairing of spin-up
fermions mediated by their $s$-wave interactions with the spin-down
particle. The lowest-order variational wave function for the trimer is
\begin{align}
  \ket{T_3(\0)} & = \sum_{\vect{k_1}\vect{k_2}}
  \gamma_{\vect{k_1}\vect{k_2}}
  c^\dag_{-\vect{k_1}-\vect{k_2}\downarrow}
  c^\dag_{\vect{k_1}\uparrow} c^\dag_{\vect{k_2}\uparrow}
  \fs.
\end{align}
Since angular and linear momentum do not commute, we restrict
ourselves to trimers with zero center-of-mass momentum ($\vect{p} =
\0$) in order to have a well-defined $L$. Such a restriction is
unlikely to be drastic since there is no physical reason to believe
that the trimer will have its lowest energy at finite momentum. In
fact, we would generally expect the energy to be higher at non-zero
$\vect{p}$ since the wave function would then contain an admixture of
higher-energy angular momentum states $L>1$.

Minimizing $\bra{T_3} (H - E) \ket{T_3}$ and defining the function
$f_{\vect{k}_2} = \sum_{\vect{k}_1} \gamma_{\k_1\k_2}$ then results in
the equation
\begin{align}\label{eq:trimerEq}
  f_{\k_2} \lbc \frac{1}{g}+ \sum_{\vect{k_1}}
  \frac{1}{E_{\vect{k_1}\vect{k_2}}} \rbc & = \sum_{\vect{k_1}}
  \frac{f_{\vect{k_1}}}{E_{\vect{k_1}\vect{k_2}}},
\end{align}
where $E_{\vect{k_1}\vect{k_2}} = -E %- 2\ef
+\epsilon_{\vect{k_1}+\vect{k_2}\downarrow} + \ekone + \ektwo$.  For
$L=1$, we have $f_\vect{k} = \tilde f_k e^{i\phi}$, where $\phi$ is
the angle with respect to the $x$-axis and $\tilde f$ is an arbitrary
function. Performing the angular integration, leaves a one-dimensional
integral equation for $\tilde f_k$ which is subsequently solved by
discretizing $k$-space and converting the integral equation into a
matrix eigenvalue equation~\cite{numerics-book}.  We emphasize that at
this order of approximation, the effect of the Fermi sea is only taken
into account in Eq.~(\ref{eq:trimerEq}) through the restriction on the
momenta in the sums ($|\k_1|>\kf$). The energy of the trimer in the
absence of a Fermi sea is recovered upon taking the limit $\kf\to0$.

\subsection{The N+1 problem}
The above approach may obviously be extended to the bound state
consisting of a single spin-$\down$ impurity and $N$ spin-$\up$
particles:
\begin{equation}
\ket{X_{N+1}}=\sum_{\k_1\ldots\k_N}\chi_{\k_1\ldots\k_N}c^\dag_{-\k_1\ldots-\k_N\down}c^\dag_{\k_1\up}\ldots
c^\dag_{\k_N\up}\fs.
\end{equation}
Again, minimizing $\bra{X_{N+1}}(H-E)\ket{X_{N+1}}$ and defining
$f_{\k_2\ldots\k_N}=\sum_{\k_1}\chi_{\k_1\ldots\k_N}$ yields an
equation for the energy of the $N+1$ bound state:
\begin{align}
&f_{\k_2\ldots\k_N}\left[\frac1g+%\frac1\Omega
\sum_{\k_1}\frac1{E_{\k_1\ldots\k_N}}\right] =
\nn \\ 
& \sum_{\k_1}
\frac{f_{\k_1\k_3\ldots\k_N}+f_{\k_2\k_1\k_4\ldots\k_N}+\ldots+f_{\k_2\ldots\k_{N-1}\k_1}}{E_{\k_1\ldots\k_N}},
\label{eq:np1}
\end{align}
where $E_{\k_1\ldots\k_N}=-E%-N\ef
+\epsilon_{\k_1+\ldots+\k_N\down}+\sum_i\eki$.  This equation is
explained in detail in Appendix \ref{app:a} where an alternative
derivation in terms of diagrams is presented.

Again the vacuum limit is recovered by letting $\kf\to0$ and we stress
that Eq.~(\ref{eq:np1}) is quite general in this limit: It is equally
valid for 1D, 2D and 3D systems; it may be extended to narrow Feshbach
resonances by letting the coupling constant be energy dependent; and
it may be used to treat the quasi-2D problem as in
Ref.~\cite{Levinsen2013}, if one includes a summation over harmonic
oscillator modes. The equation satisfied by the tetramer energy
($N=3$) in this limit was obtained for the 3D problem in
Ref.~\cite{PhysRevLett.105.223201}, the equation for the $N+1$ problem
in a quasi-2D geometry was derived in
Ref.~\cite{Levinsen2013}. Finally, Ref.~\cite{Minlos} derived an
expression similar to our Eq.~(\ref{eq:np1}) for the 3D $N+1$ vacuum
problem~\cite{notepower} and Ref.~\cite{Pricoupenko2011} generalized
this result to include also 1D and 2D.

\section{Wave functions with one particle-hole excitation\label{sec:dressed}}

We can improve on the wave functions in Sec.~\ref{sec:bare} by adding
a single particle-hole excitation on top of the Fermi sea. We expect
these improved wave functions (which are perturbative in the number of
particle-hole excitations) to provide a reasonably accurate estimate
of the single-impurity energy even in the regime of strong
interactions, $\kf\ad\sim1$, since it has been argued that
contributions from two or more particle-hole excitations nearly cancel
out via destructive interference~\cite{combescot2008}. We further
investigate the validity of this approach in Sec.~\ref{sec:2ph} where
we present our results for the impurity dressed by two particle hole
pairs.

\subsection{Polarons ($P_3$)}
Adding one particle-hole excitation to the non-interacting state
$\ket{P_1}$ gives the improved polaron wave function:
\begin{align}
\ket{P_3(\vect{p})}  & =  \alpha_{0}^{(\vect{p})} c^\dag_{\vect{p}\downarrow} \fs 
%\\ & 
+  \sum_{\vect{k}\vect{q}} \alpha_{\vect{k}\vect{q}}^{(\vect{p})} c^\dag_{\vect{p}+\vect{q} - \vect{k}\downarrow} 
c^\dag_{\vect{k}\uparrow} c_{\vect{q}\uparrow} \fs,
\label{eq:p3}
\end{align}
where we have now included a center-of-mass momentum
$\vect{p}$. Minimizing $\bra{P_3(\vect{p})}(H-E)\ket{P_3(\vect{p})}$
then gives us equations:
\begin{align}
  \lba E - \epsilon_{\vect{p}\downarrow} \rba \alpha_0^{(\vect{p})} =
  g\sum_{\vect{k}\vect{q}}
  \alpha_{\vect{k}\vect{q}}^{(\vect{p})}, \label{eq:pol0} 
\\ \notag
  \lba E - \epsilon_{\vect{p} + \vect{q} - \vect{k}\downarrow}
  - \epsilon_{\vect{k}\uparrow} + \epsilon_{\vect{q}\uparrow}
  \rba  \alpha_{\vect{k}\vect{q}}^{(\vect{p})} = & \\
  g \sum_{\vect{k'}} \alpha_{\vect{k'}\vect{q}}^{(\vect{p})} - g
  \sum_{\vect{q'}} & \alpha_{\vect{k}\vect{q'}}^{(\vect{p})} + g
  \alpha_0^{(\vect{p})}.
\label{eq:pol1}
\end{align}
We remind the reader that particle momenta $\k$ satisfy
$|\k|>\kf$. The hole momenta will be denoted by $\q$ with
$|\q|<\kf$. Combining Eqs.~(\ref{eq:pol0}) and (\ref{eq:pol1}) gives
an implicit equation for the energy,
\begin{equation}
E-\epsilon_{\p\down}=\sum_\q\left[\frac1%\Omega
g+\sum_\k
\frac{1}{E_{\p\q\k}-i0}\right]^{-1},
\label{eq:epol}
\end{equation}
with
$E_{\p\q\k}=-E+\ek-\eq+\epsilon_{\p+\q-\k\down}$. Eq.~(\ref{eq:epol})
has two solutions: The attractive and repulsive polaron which have
energies above and below the energy of the free impurity,
respectively.  Note that we must include here by hand a small
imaginary part on the right hand side in order to describe the
metastable repulsive polaron. We show in Sec.~\ref{sec:meta} how to
the extend the variational approach so that such imaginary terms arise
naturally for metastable states.

\begin{figure}
\centering
\includegraphics[width=\linewidth]{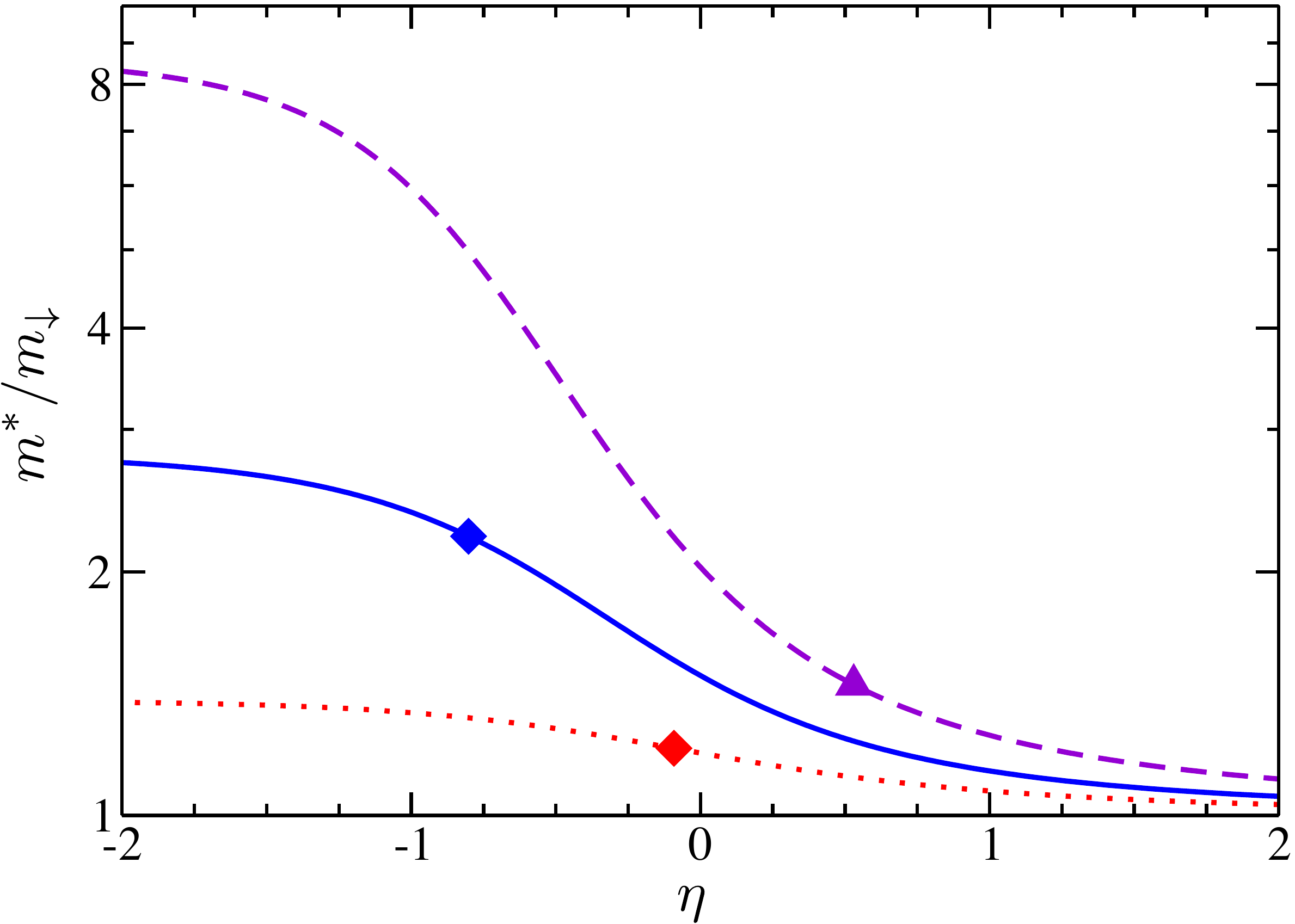}
\caption{(Color online) Effective mass of the attractive polaron
  calculated within the variational ansatz $P_3$ as a function of
  interaction $\eta \equiv \ln(\kf\ad)$ 
  for different mass ratios: $m_\up/m_\down$=6.64 (dashed), 1 (solid),
  1/6.64 (dotted). For the equal mass case, our results match those
  obtained in Ref.~\cite{Schmidt2011}. The ground state is a polaron
  for weak attractive interactions but becomes a molecule (trimer) to
  the left of the diamond (triangle). Here we compare the wavefunction
  $P_3$ with $M_4$ and $T_5$, Eqs.~(\ref{eq:m4}) and (\ref{eq:t5}).}
\label{fig:poleffmass}
\end{figure}

The solution of Eq.~(\ref{eq:epol}) at small but finite momentum
$|\p|\ll \kf$ gives the dispersion:
\begin{equation}
E(\p)=E(\0)+\frac{p^2}{2m^*},
\end{equation}
with $m^*$ the effective mass.  In Fig.~\ref{fig:poleffmass} we
display the effective mass of the attractive polaron for three
different mass ratios. Here, $r=6.64$ ($r=1/6.64$) corresponds to the
experimentally relevant lithium (potassium) impurity in a potassium
(lithium) Fermi sea. As opposed to 3D, where the effective mass was
found to diverge once the polaron was a metastable excitation
\cite{combescot2009}, we find that the effective mass is always
positive in 2D, {\em i.e.} the polaron always has its minimum energy
at $\vect{p}=0$.

In addition to the effective mass, the polarons are also described by
the wave function overlap with the free impurity --- the residue:
\begin{equation}
Z=\left|\alpha_0^{(\p)}\right|^2.
\end{equation}
The residue of the attractive polaron goes to 1 in the limit
$\kf\ad\gg1$ where the polaron is a well defined quasiparticle, while
it vanishes in the opposite limit. The repulsive polaron displays the
opposite behavior. We discuss the residue of the attractive polaron
further in Sec.~\ref{sec:2ph} where we compare with the result of
dressing the impurity by two particle-hole pair excitations.

The polaron state has been thoroughly studied in the 2D geometry. In
particular, theoretical studies of $\ket{P_3}$ have obtained the
energy of the attractive polaron \cite{Zollner:2011fk,Parish:2011vn};
the effective mass and residue of the attractive and repulsive
polarons \cite{Schmidt2011,Ngampruetikorn2012}; additionally,
Ref.~\cite{Ngampruetikorn2012} found the lowest lying repulsive
polaron state in the limit $\kf\ad\gg1$ to have finite momentum.
Experimental evidence of both the repulsive and the attractive polaron
was found using radiofrequency spectroscopy in
Ref.~\cite{Koschorreck2012}.

\subsection{Molecules ($M_4$)}
We can likewise improve on the bare molecule state $\ket{M_2}$ by
adding one particle-hole pair as follows:
\begin{align}\notag
  \ket{M_4(\vect{p})} & = \sum_{\vect{k}}
  \varphi_{\vect{k}}^{(\vect{p})} c^\dag_{\vect{p}-\vect{k}\downarrow}
  c^\dag_{\vect{k}\uparrow} \fs \\ \label{eq:m4} & +
  \sum_{\vect{k}\vect{k'}\vect{q}}
  \varphi_{\vect{k}\vect{k'}\vect{q}}^{(\vect{p})} c^\dag_{\vect{p} +
    \vect{q}-\vect{k} - \vect{k'}\downarrow} c^\dag_{\vect{k}\uparrow}
  c^\dag_{\vect{k'}\uparrow} c_{\vect{q}\uparrow} \fs.
\end{align}
The minimization procedure now leads to the following equations for
the energy of the molecule:
\begin{align}
\lbc \frac1g+%\frac1\Omega
\sum_{\k}\frac1{E_{\p\k}}\rbc & =
-\sum_{\k\q}\frac{G_{\q\k}}{E_{\p\k}},
\label{eq:mol1}
\\
%&&\hspace{-10mm}
\lbc \frac1g+%\frac1\Omega
\sum_{\k}\frac1{E_{\p\q\k_1\k}}\rbc
G_{\q\k_1} & = \nn \\ 
%&&\hspace{-2mm}
 %\frac1\Omega
\sum_{\k}\frac{G_{\q\k}}{E_{\p\q\k_1\k}} &
-\frac1{E_{\p\k_1}}
-%\frac1\Omega
\sum_{\q'}\frac{G_{\q'\k_1}}{E_{\p\k_1}},
\label{eq:mol2}
\end{align}
where $E_{\p\k}=-E%-\ef
+\ek+\epsilon_{\p+\k\down}$ and $E_{\p\q\k_1\k_2}=-E%-\ef
+\ekone+\ektwo- \eq+\epsilon_{\p+\q-\k_1-\k_2\down}$. The function $G$
is defined as
$G_{\q\k}\equiv\sum_{\k'}\varphi^{(\p)}_{\k\k'\q}/\sum_{\k'}\varphi^{(\p)}_{\k'}$.
The energy of the molecule in the 2D geometry was first obtained in
Ref.~\cite{Parish:2011vn}. The coupled integral equations
(\ref{eq:mol1}) and (\ref{eq:mol2}) are equivalent to the equation for
the molecule energy obtained in
Refs.~\cite{combescot2009,mora2009,punk2009} for an impurity in a 3D
Fermi gas.

\begin{figure}
\centering
\includegraphics[width=\linewidth]{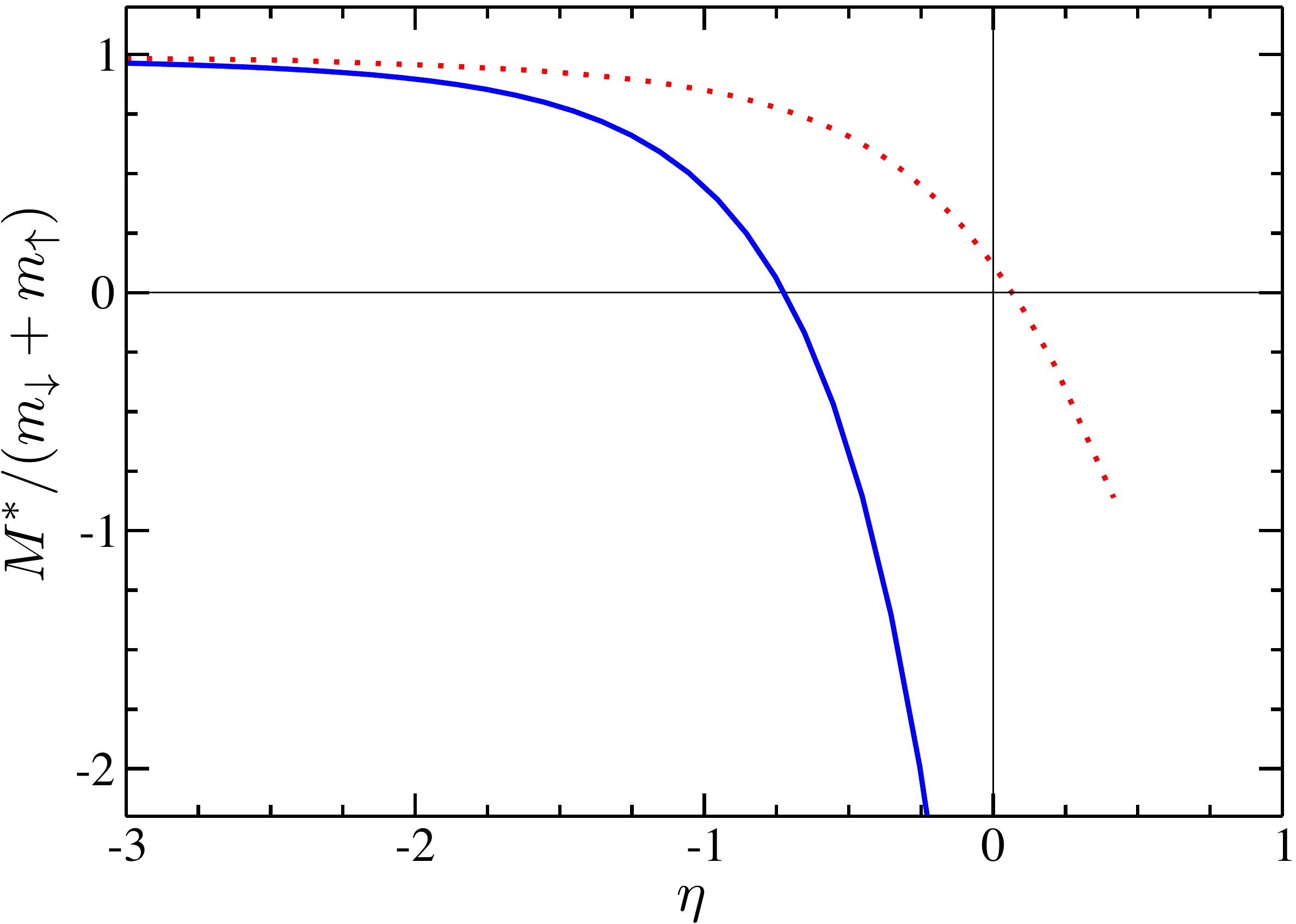}
\caption{(Color online) Inverse effective mass of the molecule as a function
  of interaction parameter for different mass ratios:
  $m_\up/m_\down$= 1 (solid), 1/6.64 (dotted).}
\label{fig:effmass}
\end{figure}

\begin{figure}
\centering
\includegraphics[width=\linewidth]{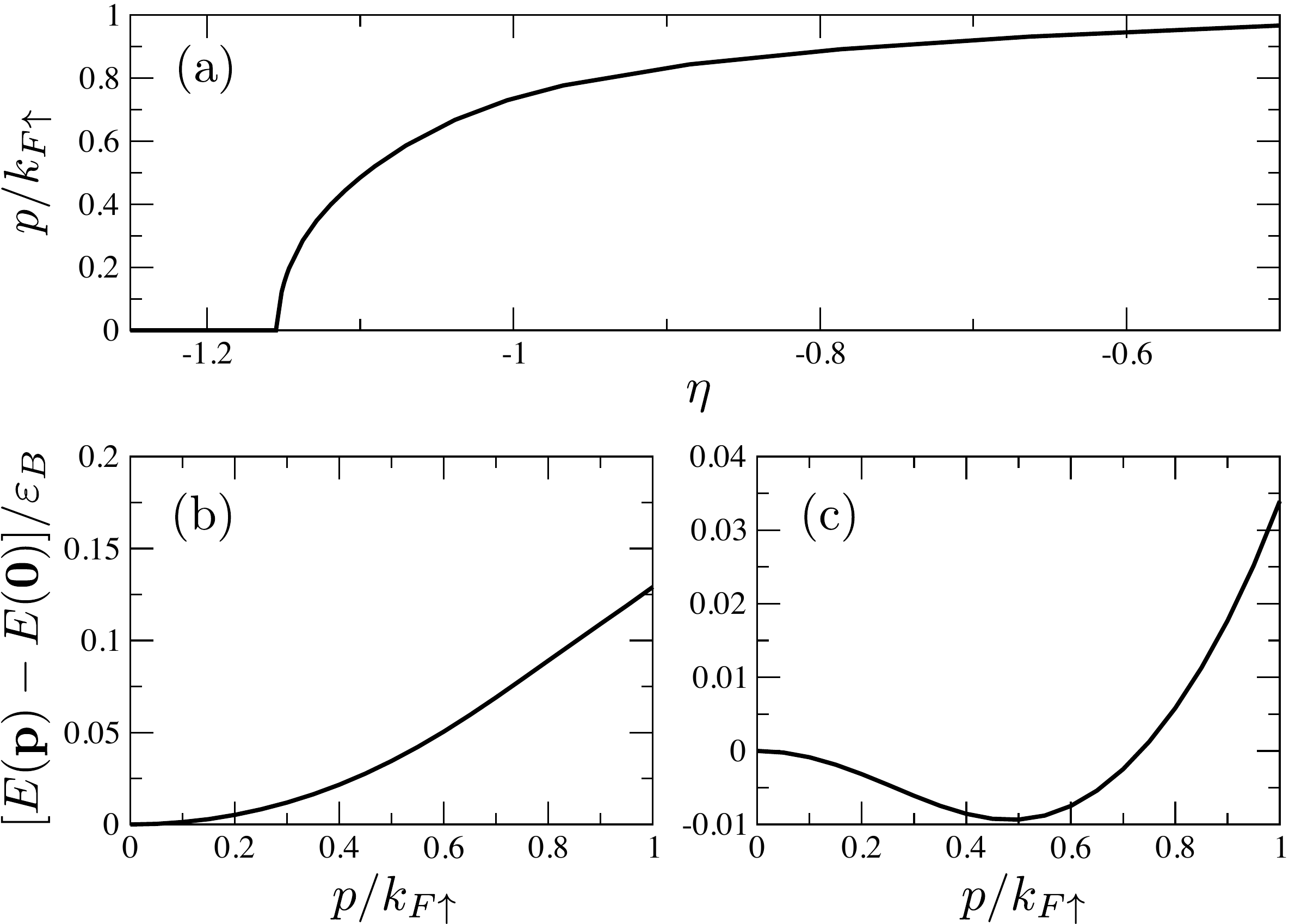}
\caption{Illustration of the dispersion of $M_4(\p)$ for $m_\up=2m_\down$. (a)
  Momentum at the minimum of the dispersion as a function of
  interaction parameter. (b) Dispersion at $\eta=-1.25$ and (c)
  dispersion at $\eta=-1.1$.}
\label{fig:molfinitep}
\end{figure}

Similarly to the polaron above, the energy of the molecule as a function of
momentum yields the dispersion
\begin{equation}
E(\p)=E(\0)+\frac{p^2}{2M^*},
\end{equation}
with $M^*$ the effective mass. In the absence of interactions, the
effective mass of the molecule is simply $M^*=m_\up+m_\down$. However,
we observe a strong dependence of the effective mass on the
interaction parameter, as shown in Fig.~\ref{fig:effmass}. As further
illustrated in Fig.~\ref{fig:molfinitep}, once $M^*<0$ the molecule
has its minimum energy at finite momentum, {\em i.e.} the pairing
occurs at a finite momentum. As mentioned previously, this corresponds
to the FFLO phase in the limit of large polarization, and we find that
this phase occupies regions of the phase diagrams discussed in
Sec.~\ref{sec:phase}.

\subsection{Trimers  ($T_5$)}

As discussed above, in the absence of a Fermi sea and for $r > 3.33$,
the impurity binds two particles to form a trimer. Since we wish to
investigate the possibility of the trimer being the ground state
across the regime of strong interactions, we use the dressed wave
function
\begin{align} \notag &\ket{T_5(\0)} = \sum_{\vect{k_1}\vect{k_2}}
  \gamma_{\vect{k_1}\vect{k_2}}
  c^\dag_{-\vect{k_1}-\vect{k_2}\downarrow} c^\dag_{\vect{k_1}\uparrow} c^\dag_{\vect{k_2}\uparrow} \fs \\
  & +\sum_{\vect{k_1}\vect{k_2}\vect{k}\vect{q}}
  \gamma_{\vect{k_1}\vect{k_2}\vect{k}\vect{q}}
  c^\dag_{\vect{q}-\vect{k_1}-\vect{k_2}-\vect{k}\downarrow}
  c^\dag_{\vect{k_1}\uparrow} c^\dag_{\vect{k_2}\uparrow}
  c^\dag_{\vect{k}\uparrow} c_{\vect{q}\uparrow} \fs.
\label{eq:t5}
\end{align}
Following the minimization procedure, we find two coupled integral
equations:
\begin{align}
%&&\hspace{-10mm}
  J_\vect{k_1}  \left[ \frac{1}{g} + 
\sum_{\vect{k_2}}
    \frac{1}{E_{\vect{k_1}\vect{k_2}}}\right] = 
\sum_{\k_2}\frac{J_\vect{k_2}}{E_{\vect{k_1}\vect{k_2}}}  -
\sum_{\vect{q}\k_2}\frac{G_{\q\k_1\k_2}}{E_{\vect{k_1}\vect{k_2}}} & ,
\label{eq:trima} \\
%&& \hspace{-10mm}
 G_{\q\k_1\k_2} \left[\frac1g  +
\sum_{\k} \frac1{E_{\q\k_1\k_2\k}}\right]
=-\frac{J_\vect{k_1}-J_\vect{k_2}}{E_{\vect{k_1}\vect{k_2}}} & 
 \nn \\
% && \hspace{4mm}
 +\sum_{\vect{k}} \frac{G_{\q\vect{k_1}\vect{k}} + G_{\q\vect{k}
\vect{k_2}}}{E_{\q\vect{k_1}\vect{k_2}\vect{k}}}
-\sum_{\vect{q'}}\frac{G_{\q'\k_1\k_2}}{E_{\vect{k_1}\vect{k_2}}} & .\label{eq:trimb}
\end{align}
We have defined $J_\k = g\sum_{\vect{k'}} \gamma_{\vect{k}\vect{k'}}$
and $G_{\q\vect{k_1}\vect{k_2}} = 3 g \sum_{\vect{k'}}
\gamma_{\vect{k_1}\vect{k_2}\vect{k'}\q}$. The energies are
$E_{\vect{k_1}\vect{k_2}} = -E %-2\ef+ \epsilon_{\vect{k_1}\uparrow} +
%\epsilon_{\vect{k_2}\uparrow}
+\ekone+\ektwo
+\epsilon_{\vect{k_1} +
  \vect{k_2}\downarrow} $ and $E_{\q\vect{k_1}\vect{k_2}\vect{k}} = -E%-2\ef
%+ \epsilon_{\vect{k_1}\uparrow} + \epsilon_{\vect{k_2}\uparrow} +
%\epsilon_{\vect{k}\uparrow}
+\ekone+\ektwo+\ek
-\eq+ \epsilon_{\q-\vect{k} - \vect{k_1} -
  \vect{k_2}\downarrow}$.  Eqs.~(\ref{eq:trima}) and (\ref{eq:trimb})
were first derived and solved for the trimer energy in
Ref.~\cite{Mathy:2011ys} for the impurity problem in a 3D Fermi
gas. The projection onto the $p$-wave trimer state is performed by
taking
\begin{equation}
f_{\k_1}= \tilde f_{k_1}e^{i\phi_1},\hspace{4mm}
G_{\q\k_1\k_2}=\tilde G(q,k_1,k_2,\Delta\phi_{1q},\Delta\phi_{2q})e^{i\phi_1},
\label{eq:pwavetrim}
\end{equation}
with $\phi_1$, $\phi_2$, and $\phi_q$ the angles which $\k_1$, $\k_2$,
and $\q$ make with the axis of reference, while
$\Delta\phi_{1q}=\phi_1-\phi_q$ and
$\Delta\phi_{2q}=\phi_2-\phi_q$. 

\subsection{The $N+1$ problem}

Like the bare wavefunctions of Sec.~\ref{sec:bare}, the above approach
may be extended to the study of the bound states of the impurity and
$N$ spin-$\up$ fermions. The variational wave function is dressed by
one particle-hole pair excitation of the Fermi sea and the
minimization procedure carried out as above. This leads to two coupled
integral equations similar to Eqs.~(\ref{eq:trima}) and
(\ref{eq:trimb}) for the trimer above. The equations are derived in
Appendix~\ref{app:b} using the diagrammatic technique. We shall not
attempt here to solve for the energy of the tetramer or bound states
containing even more particles.

\section{Phase diagrams \label{sec:phase}}

\begin{figure}
\centering
\includegraphics[width=\linewidth]{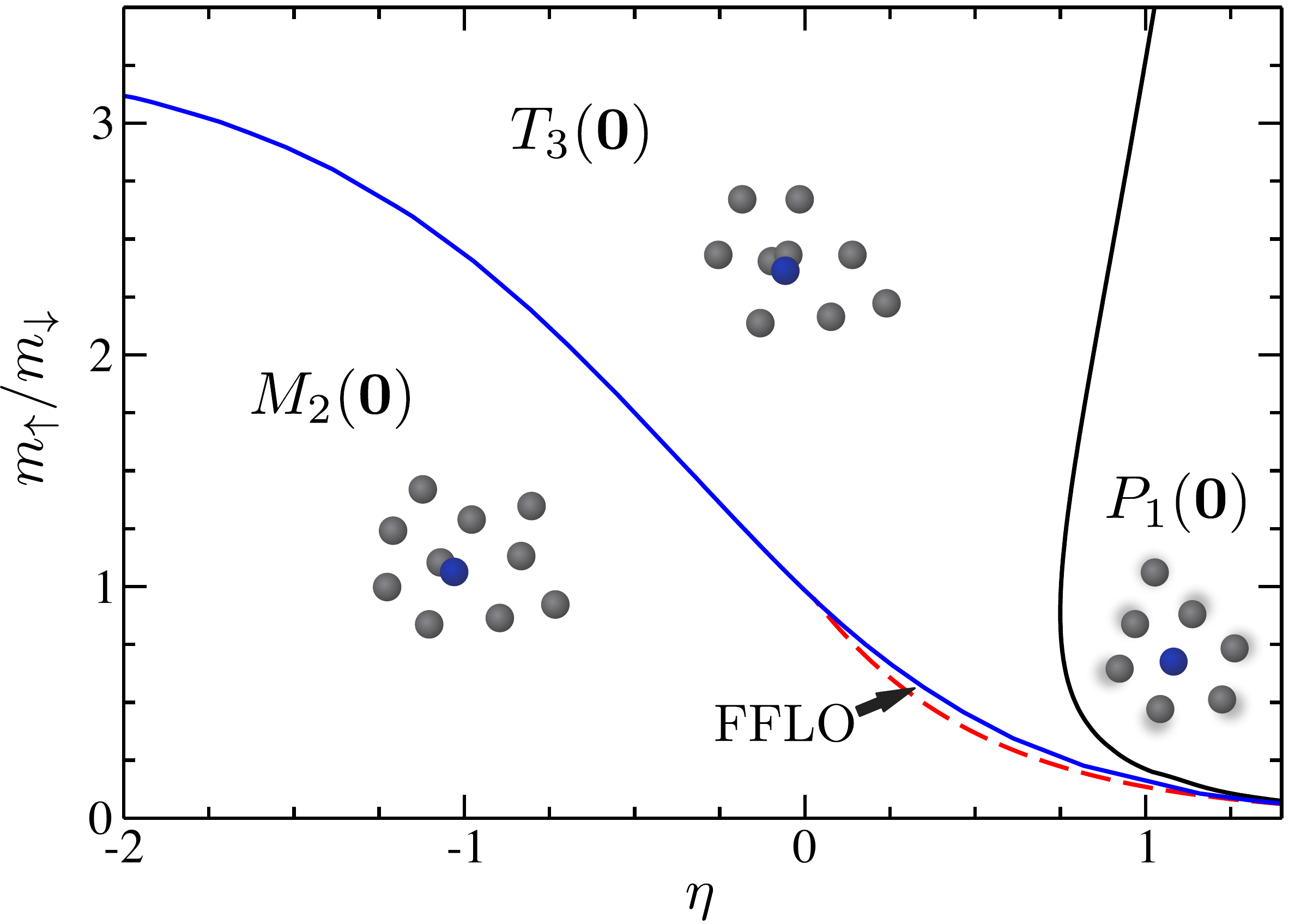}
\caption{(Color online) Ground state phase diagram for an impurity
  atom attractively interacting with a 2D Fermi gas.  The FFLO phase
  corresponds to the molecule $M_2(\vect{p})$ with non-zero momentum
  $\vect{p}$ in the ground state. The behavior of the momentum at
  which the energy is at its minimum is given by
  Eq.~\eqref{eq:M2_momentum} and it goes smoothly to zero at the
  dashed (red) line given by $m_\up/m_\down = 1/(k_{F}\ad)^2$.  We
  find that small slivers of FFLO and trimer phases remain as $\eta
  \to \infty$.  Note, also, that the trimer exists above the critical
  mass ratio $r \simeq 3.33$ in the limit $\eta \to -\infty$, which
  agrees with the result for the 3-body bound state in a
  vacuum~\cite{2Dtrimer}.  }
\label{fig:2DFermi}
\end{figure}

We now determine the ground state for the single impurity and the
correponding binding transitions.  In Fig.~\ref{fig:2DFermi} we show
the phase diagram for the ``undressed'' wave functions of
Sec.~\ref{sec:bare}.  Surprisingly, we find that a molecule existing
at a given mass ratio $r<3.33$ must always first bind an extra spin-up
fermion to form a trimer before it can unbind into a polaron. This
appears to be an artifact of the approximation. However, it does
signify the importance of three-body correlations for all mass ratios
in 2D. Additionally, we find a sliver of FFLO phase, corresponding to
a finite momentum molecule, on the border of the zero-momentum
molecule and the trimer phases.  The large region of trimer phase
below $r\simeq3.33$ appears to result from the fact that the FFLO
molecule is unstable towards binding an extra $\up$ particle, like in
3D~\cite{Mathy:2011ys}.  Whereas trimers are favored by the medium, we
find that tetramers consisting of three spin-$\up$ particles and the
impurity appear to be disfavored, {\em i.e.} the phase transition is
found to occur at larger mass ratios in the medium than the critical
mass ratio of $r=5.0$ in vacuum~\cite{Levinsen2012}. This suggests
that four-body correlations are not as important in the many-body
system at low mass ratios as might be initially expected.

\begin{figure}
\centering
\includegraphics[width=\linewidth]{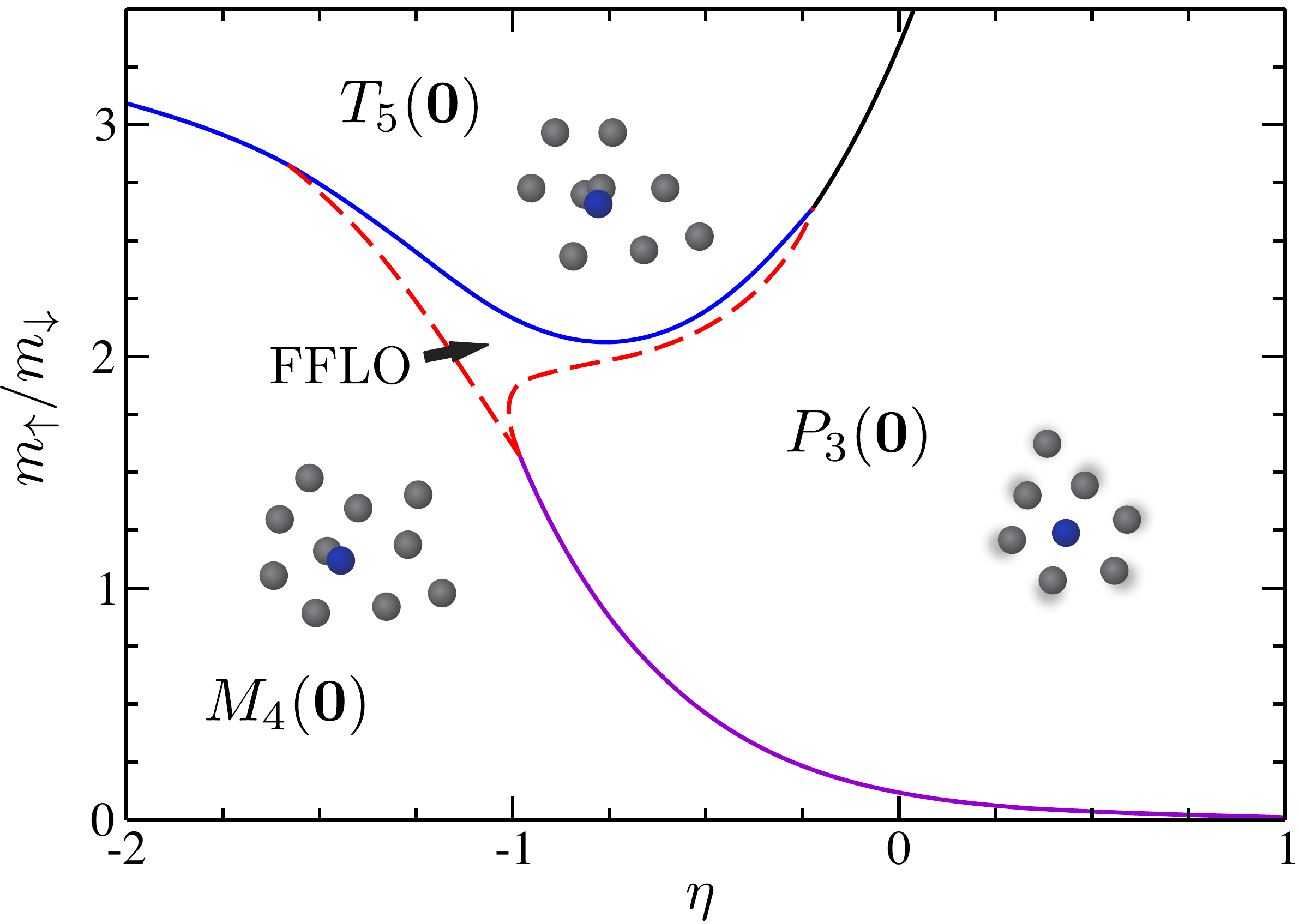}
\caption{(Color online) Ground state phase diagram for a spin-down
  impurity atom immersed in a 2D Fermi gas, with phase boundaries
  calculated using the dressed wave functions of
  Sec.~\ref{sec:dressed}.  The FFLO phase corresponds to the molecule
  $M_4(\vect{p})$ with non-zero momentum $\vect{p}$ in the ground
  state. See Fig.~\ref{fig:molfinitep} for the behavior of $\vect{p}$
  as a function of $\eta$ across the FFLO region.}
\label{fig:2DFermi2}
\end{figure}

Next, Fig.~\ref{fig:2DFermi2} shows our phase diagram obtained using
wave functions dressed by one particle-hole pair -- see
Sec.~\ref{sec:dressed}. As above, we find that the trimer is favored
by the Fermi sea, but it now does not appear below a mass ratio of
$r\approx2.1$. Additionally, we find that the FFLO region, where the
ground-state molecule has finite momentum, is enlarged at this level
of approximation, making it possible that the FFLO phase may be
observed in this system.

We expect this phase diagram to be qualitatively correct also across
the regime of strong many-body corrections, $|1/\eta|\ll1$, as
contributions from two or more particle-hole pairs cancel
approximately~\cite{combescot2008} and parts of the phase diagram are
fixed by perturbative and exact calculations. In the limit of large
negative $\eta$, the dressing of the molecule and trimer by one
particle-hole pair yields the correct form of the first-order
correction to their energy due to the interaction with the Fermi sea
in a perturbative expansion in $1/|\eta|$. Thus the phase transition
from $M_4(\0)$ to $T_5(\0)$ and the fact that the trimer is favored by
the Fermi sea is quantitatively robust. Furthermore, for an impurity
with infinite mass ($r=0$), the approach correctly identifies the
dimer as the ground state~\cite{Parish:2011vn}.

\section{Accuracy of the variational approach \label{sec:2ph}}

\begin{figure}
\centering
\includegraphics[width=\linewidth]{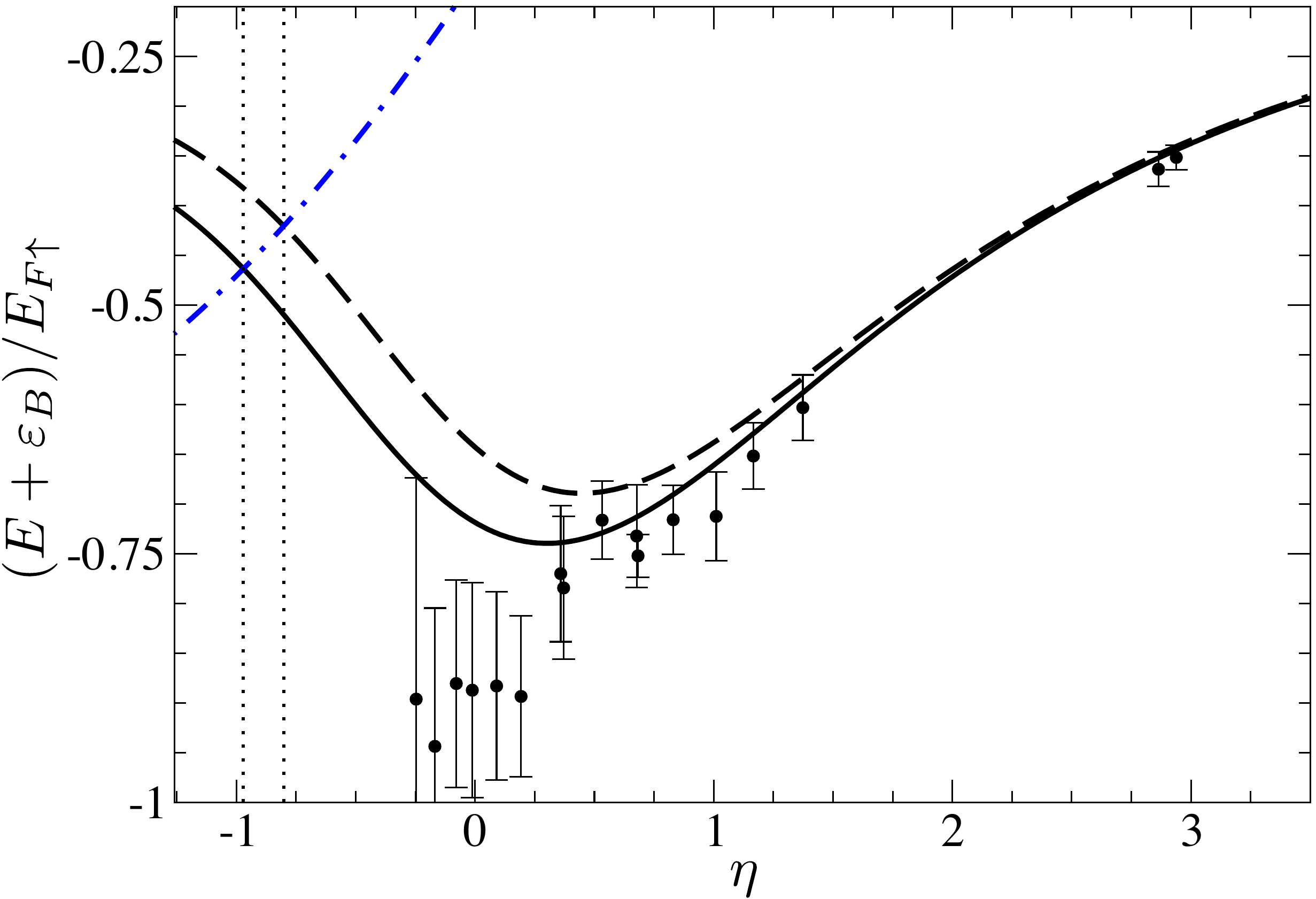}
\caption{(Color online) Energy measured from the two-body binding
  energy. The dashed line corresponds to the results for the ansatz
  $\ket{P_3}$, Eq.~(\ref{eq:p3}), the solid line to $\ket{P_5}$,
  Eq.~(\ref{eq:p5}), and the dot-dashed line to the molecule ansatz
  $\ket{M_4}$, Eq.~(\ref{eq:m4}). The polaron-molecule transitions in
  the two approximations are illustrated by vertical dotted lines,
  while the experimental data is taken from
  Ref.~\cite{Koschorreck2012}, rescaled to reflect the quasi-2D nature
  of the experiment \cite{noterescale}.}
\label{fig:2ph}
\end{figure}

The variational approach employed above has been validated by several
methods in various settings. In 3D, the energy and effective mass of
the attractive polaron quasiparticle have been measured in
Refs.~\cite{PhysRevLett.102.230402} and~\cite{PhysRevLett.103.170402},
respectively, and good agreement with the variational
calculation~\cite{chevy2006_2, Combescot:2007bh} was obtained, even
close to the unitary limit. It has also been shown that the polaron
and molecule
energies~\cite{combescot2008,combescot2009,mora2009,punk2009}
calculated using this method for a 3D Fermi gas with $m_\uparrow =
m_\downarrow$ are in good agreement with those from quantum Monte
Carlo~\cite{prokofiev2008,prokofiev2008_2}. More recently, an
experimental and theoretical investigation of the population
imbalanced $^{40}$K-$^6$Li mixture showed impressive agreement between
the measured energies, residues, and lifetimes of the attractive and
repulsive polarons when compared with the theoretical predictions from
the diagrammatic technique~\cite{innsbruck} (which is equivalent to
the variational one presented here). Finally, it has been argued that
contributions from two or more particle-hole excitations nearly cancel
out via destructive interference~\cite{combescot2008}. This argument
does not depend on the dimension and indeed in the one-dimensional
geometry it was shown that the variational ansatz agrees well with the
exact Bethe Ansatz solution \cite{1dpolaron}.

We now wish to demonstrate explicitly the accuracy of the dressed wave
functions of the previous section. We emphasize the perturbative
nature of the many-body system as long as $1/|\ln(\kf\ad)|\ll1$, but
we wish here to quantify the effect of quantum fluctuations in the
strongly interacting region, $\kf\ad\sim1$. To this end we write down
a variational wave function for the impurity dressed by two
particle-hole pairs:
\begin{eqnarray}
\ket{P_5(\0)} & = & \alpha_{0} c^\dag_{\vect{0}\downarrow} \fs 
+  \sum_{\vect{k}\vect{q}} \alpha_{\vect{k}\vect{q}} c^\dag_{\vect{q} - \vect{k}\downarrow} 
c^\dag_{\vect{k}\uparrow} c_{\vect{q}\uparrow} \fs
\nn \\ && \hspace{-16mm}
+  \sum_{\tiny\begin{array}{c}\k_1\k_2 \\\q_1\q_2\end{array}} \alpha_{\vect{k_1}\k_2\q_1\vect{q_2}} c^\dag_{\vect{q_1}+\q_2 - \vect{k_1}-\k_2\downarrow} 
c^\dag_{\vect{k_1}\uparrow} c^\dag_{\vect{k_2}\uparrow}
c_{\vect{q_1}\uparrow} 
c_{\vect{q_2}\uparrow}
\fs.\nn\\ &&
\label{eq:p5}
\end{eqnarray}
This wave function was first studied for the 3D polaron problem in
Ref.~\cite{combescot2008}. The minimization of
$\bra{P_5}(H-E)\ket{P_5}$ then results in two coupled integral
equations,
\begin{eqnarray}
\hspace{-10mm}
f_\q\lbc\frac1g+\sum_\k\frac1{E_{\q\k}}\rbc & = &
\sum_{\q'}\frac{f_{\q'}}{E}-\sum_{\k\q'}
\frac{G_{\q\q' \k}}{E_{\q\k}},
\label{eq:p5a}
\\ &&
\hspace{-35mm}
G_{\q_1\q_2\k}\lbc\frac1g+\sum_{\k'}\frac1{E_{\q_1\q_2\k\k'}}\rbc
=\sum_{\k'}\frac{G_{\q_1\q_2\k'}}{E_{\q_1\q_2\k\k'}} \nn \\ &&
\hspace{-25mm}
-\frac{f_{\q_1}+\sum_{\q_2'}G_{\q_1\q_2'\k}}{E_{\q_1\k}}
+\frac{f_{\q_2}-\sum_{\q_1'}G_{\q_1'\q_2\k}}{E_{\q_2\k}}.
\label{eq:p5b}
\end{eqnarray}
The energies are $E_{\q\k}=-E+\ek-\eq+\epsilon_{\q-\k\down}$ and
$E_{\q_1\q_2\k_1\k_2}=-E+\ekone \ektwo-\eqone-\eqtwo
+\epsilon_{\q_1+\q_2-\k_1-\k_2\down}$, and we define
$G_{\q_1\q_2\k}=4g\sum_{\k'} \alpha_{\k'\k\q_1\q_2}$ and
$f_\q=g\sum_\k\alpha_{\k\q}$. The restriction to $s$-wave scattering
of the impurity off a majority atom means that $f_\q$ depends only on
the magnitude of $\q$. Likewise, the vertex $G$ depends on the
magnitudes of $\q_1$, $\q_2$, and $\k$ and the two angles
$\angle_{\q_1,\k}$ and $\angle_{\q_2,\k}$.

Our results for the polaron energy for equal masses of the two species
are shown in Fig.~\ref{fig:2ph}. We see that the two ans\"atze agree
very well in the weakly interacting regime, $\eta\gg1$. In the regime
of strong many-body effects, $1/|\eta|\ll1$, the polaron energies
resulting from the two ans\"atze show the same non-monotonic behavior
and are never further separated than by 0.1$\ef$. The polaron in a
quasi-2D geometry was observed in a recent
experiment~\cite{Koschorreck2012}. The data from the experiment is
also displayed in Fig.~\ref{fig:2ph} and is seen to match the energy
of the $\ket{P_5}$ ansatz quite well for $\eta\gtrsim0.25$, leading us
to believe that the $\ket{P_5}$ ansatz is likely to closely reproduce
the true energy of the polaron. The reason for the discrepancy below
this value may be due to the finite interaction range of the
interatomic potential, a finite density of impurity atoms, finite
temperature effects, and trap-averaging effects.

From the polaron energies in the two ans\"atze we obtain the following
values of the interaction parameter at the polaron-molecule
transition:
\begin{align}
P_3\mbox{-}M_4 \mbox{ transition: }  & \ \eta=-0.80, \,\,
\eb/\ef=9.9 \nn \\
P_5\mbox{-}M_4 \mbox{ transition: } & \ \eta=-0.97, \,\, \eb/\ef=14.0
%\nn \\ 
\label{eq:pmtrans}
\end{align}
This compares well with the experimental result of Koschorreck {\em et
  al.}~\cite{Koehl2012}, $\eta=-0.88(0.20)$~\cite{noterescale}. 

If the polaron and molecule variational wavefunctions converge with
increasing numbers of particle-hole pair excitations, then it is
possible to show that the above results for the polaron-molecule
transition, Eq.~\eqref{eq:pmtrans}, provide upper and lower bounds for
the actual polaron-molecule transition.  First, we define
$E_{2n-1}-\eb$ and $E_{2n}-\eb$ to be the energies for the polaron
$\ket{P_{2n-1}}$ and molecule $\ket{M_{2n}}$, respectively, with
$n\geq 1$. Note that we must have $E_{2n+1}\leq E_{2n-1}$ and
$E_{2n+2}\leq E_{2n}$ since larger $n$ corresponds to successively
better variational wave functions with successively lower
energies. Now we assume for each $\eta$ that the corrections to the
energies become smaller with increasing $n$ such that
\begin{align}
 E_{2n+1} - E_{2n+3} \leq E_{2n} - E_{2n+2} \leq E_{2n-1} - E_{2n+1}
\end{align}
and we also assume that $E_{2n-1}$ ($E_{2n}$) is a monotonically
decreasing (increasing) function of $\eta$ around the transition like
in Fig.~\ref{fig:2ph}, i.e.\ near a given point $\eta_0$ we can write
\begin{align}\nn
E_{2n} (\eta) & = E_{2n}(\eta_0) + \rho_M (\eta-\eta_0) \\ \nn
E_{2n-1}(\eta) & =  E_{2n-1}(\eta_0) - \rho_P (\eta-\eta_0)
\end{align}
where $\rho_P, \rho_M > 0$. Then for successively better
approximations for the polaron-molecule transition:
\begin{align} \nn
E_{2n}(\eta^+) & = E_{2n-1}(\eta^+) \\ \nn
E_{2n}(\eta^-) & = E_{2n+1}(\eta^-) \\ \nn
E_{2n+2}(\eta^\prime) & = E_{2n+1}(\eta^\prime)
\end{align}
one can prove that $\eta^- < \eta^\prime < \eta^+$.  Therefore the
$P_5$-$M_6$ transition will lie between the $P_5$-$M_4$ and
$P_3$-$M_4$ transitions, and so on, thus proving that the results
\eqref{eq:pmtrans} provide upper and lower bounds.

\begin{figure}
\centering
\includegraphics[width=\linewidth]{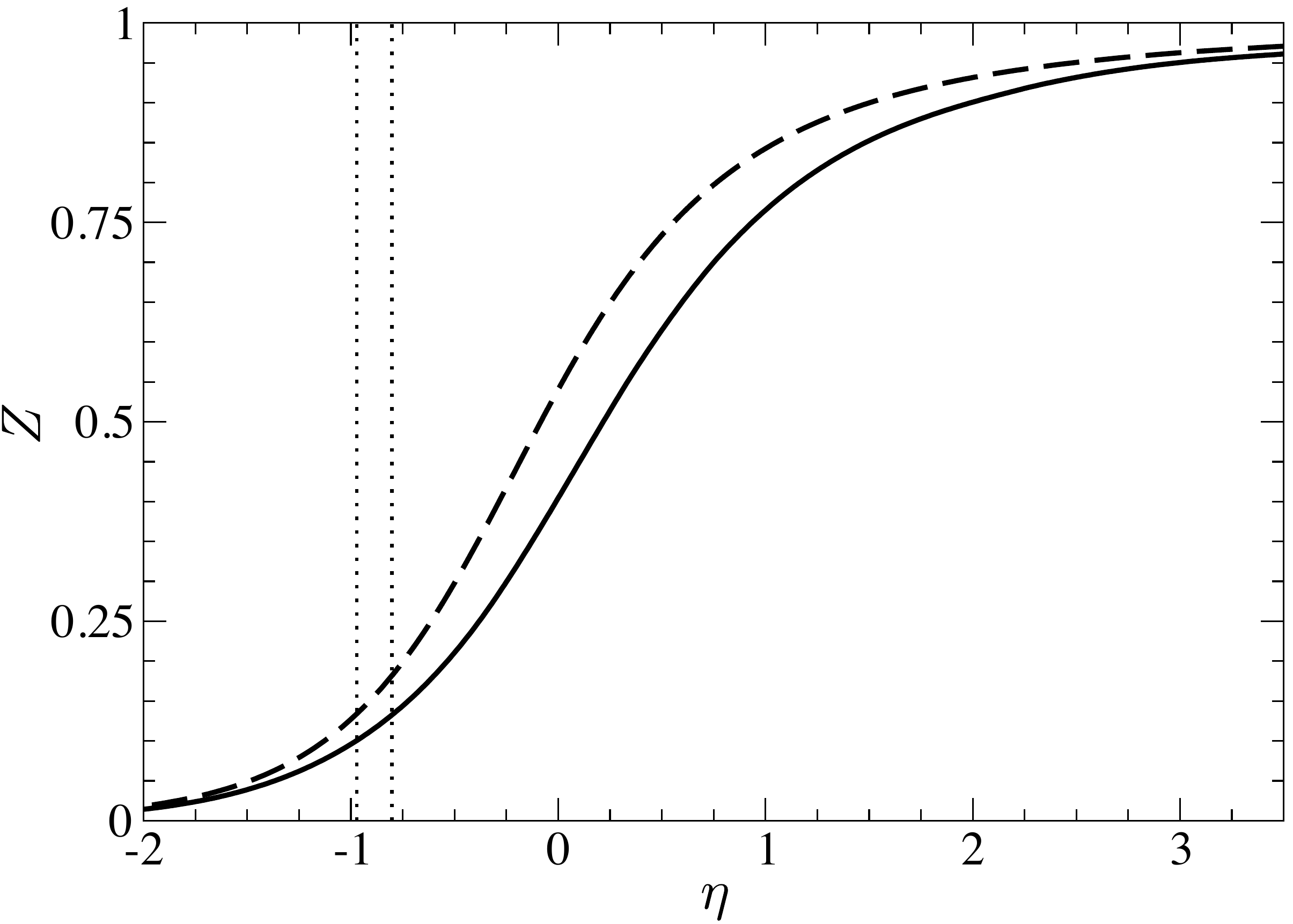}
\caption{Residue of the attractive polaron for equal masses calculated
  from the variational wavefunctions $P_5(\0)$ (solid line) and
  $P_3(\0)$ (dashed). The dashed vertical
  lines represent the
  polaron-molecule transitions calculated within the two
  approximations.
  \label{fig:2phres}}
\end{figure}

Finally, we also calculate the wavefunction overlap with the free
impurity, the residue $Z$.  As illustrated in Fig.~\ref{fig:2phres},
the residue also agrees quite well between the two ans\"atze, even
across the regime of strong many-body effects.  Note that the residue
is finite at the polaron-molecule transition, indicating that the
residue jumps to zero discontinously when the impurity binds an extra
$\up$ particle. This is a signature of a sharp ``first-order'' binding
transition like in 3D~\cite{punk2009,bruun2010}.  To our knowledge,
our result for $\ket{P_5}$ is the best estimate for the polaran
residue thus far presented in the literature.

\section{Finite density \label{sec:dens}}
The high polarization limit of the phase diagram for the
spin-imbalanced Fermi gas involves a finite density of impurities, and
thus the question arises whether the single-impurity transitions are
\emph{thermodynamically} stable, i.e., whether they are preempted by
first-order transitions in the thermodynamic limit. Typically, one
requires the interactions between dressed impurities in order to
assess this scenario; for instance, an attractive interaction between
bosonic dressed impurities (e.g.\ dressed dimers or tetramers for a
spin-down fermionic impurity atom) implies that the dressed impurities
are unstable towards collapse into a region of higher density.
However, the approach we will employ here is to determine whether the
conditions for phase separation between the superfluid (SF) and
fully-polarized normal (N) phases are ever satisfied. To this end, we
assume that the superfluid is unpolarized and exploit the equal-mass
equation of state for the BCS-BEC crossover derived from QMC
calculations~\cite{PhysRevLett.106.110403}. This assumption is
reasonable since no polarized superfluid is observed in the mean-field
calculations for the 2D Fermi gas~\cite{conduit2008}.

The onset of phase separation in a Fermi gas
near full polarization corresponds to the following
conditions for the pressures and
chemical potentials in each phase:
\begin{align}\label{eq:pressure}
p_{SF} & = p_{N}\,, \ \ \mu^{SF}_\sigma = \mu^{N}_\sigma \,, \ \
\mu_\downarrow^{N} = E \,,
\end{align}
where $E$ is the energy of a single $\down$ impurity immersed in a
spin-polarized Fermi gas.
At the mean-field level, this is very easily carried out by setting $E
= 0$, i.e.\ assuming that the impurity atom is non-interacting, and
then analyzing the minima of the thermodynamic potential
$\Omega(\Delta,\mu_\sigma) \equiv -pV$, where $\Delta$ is the
mean-field superfluid order parameter. Identifying the points at which
we have degenerate minima at $\Delta = 0$ and $\Delta \neq 0$
corresponds exactly to satisfying the conditions
\eqref{eq:pressure}. Using the mean-field approach, we obtain a
first-order transition at $\eb/\ef = \sqrt{1+r}-1$.  Below, we obtain
a more accurate result for the case of equal masses $m_\up = m_\down
\equiv m$.

Clearly, the pressure and spin-up chemical potential in the
fully-polarized normal phase are known exactly:
\begin{align}
p_N & = \frac{k_{F\uparrow}^4}{16\pi m}  \\
\mu^N_\uparrow & = \ \ef = \frac{k_{F\uparrow}^2}{2m}
\end{align}
Thus, we just need to estimate the binding energy $E \equiv A \ef$ to
determine $\mu^N_\downarrow$.  We obtain estimates from both wave
functions $\ket{P_3}$ and $\ket{P_5}$.

The pressure and average chemical potential for the unpolarized
superfluid can be written:
\begin{align}
p_{SF} & = \frac{k_{F}^4}{8\pi m}  \zeta_p(\nu) \\
\mu^{SF} & = \frac{\mu^{SF}_\up + \mu^{SF}_\down}{2} = \frac{k_{F}^2}{2m} \zeta_\mu (\nu)
\end{align}
where $\zeta_p(\nu)$, $\zeta_\mu(\nu)$ are interpolating functions
that can be determined from the QMC equation of
state~\cite{PhysRevLett.106.110403}, and $\nu = \ln(k_F \ad) = \eta
+\ln(k_F/\kf)$.  The Fermi momentum $k_F$ of the superfluid is
generally different from that in the fully-polarized phase.

By equating pressures and chemical potentials in each phase, we obtain
the coupled equations
\begin{align}
\left( \frac{\kf}{k_F} \right)^4  = 2 \zeta_p(\nu),  \ \ & \ \ 
\frac{\kf}{k_F}  = \sqrt{\frac{2 \zeta_\mu(\nu)}{1+A}}
\end{align}
which must be solved for $\eta$ and $\kf/k_F$.  For the SF-N
transition, we thus get within the two approximations for the impurity
wave function
\begin{align} \notag
\ket{P_3}:  \  \ & \ \eta = -0.69, \ \  \eb/\ef = 7.9 \\ \label{eq:firstorder}
\ket{P_5}:  \  \ & \ \eta = -0.90, \ \  \eb/\ef = 12.2
\end{align}
By comparing these values with \eqref{eq:pmtrans}, we see that this
first-order transition (and the concomitant phase separation) preempts
the single-impurity transition in the thermodynamic limit.  This
implies that the single-impurity transition effectively corresponds to
a spinodal line for the SF-N transition, and this could be the case
for much of the polaron-molecule transition line in
Fig.~\ref{fig:2DFermi2}.  Note, however, that the results
\eqref{eq:firstorder} strictly apply to zero temperature and require
the existence of a superfluid. Such a SF-N transition may be destroyed
by strong thermal fluctuations and so we can envisage a scenario where
we only have the single-impurity transition existing at finite
temperature.  Future work is required to distinguish between these
possible scenarios.

\section{Variational description of metastable states \label{sec:meta}}
The variational approach can also be generalized to study metastable
excited states such as the repulsive
polaron~\cite{Schmidt2011,Ngampruetikorn2012,Cui:2010zr,Massignan:2011fk}.
Rather than minimizing the energy of the variational wave function,
one instead needs to construct its equations of motion by minimizing
the ``error'' quantity~\cite{McLachlan64,PhysRevE.51.5688}
\begin{align}\notag
 \int \bra{\psi(t)} \delta^\dag \delta \ket{\psi(t)} dv
\end{align}
for all allowed variations of the unknown function $i %\hbar
\del_t\psi(t)$, where $\delta = i \del_t - H$.  For the polaron wave
function \eqref{eq:p3}, this yields the same equations \eqref{eq:pol0}
and \eqref{eq:pol1}, but with energy $E$ replaced with $i %\hbar
\del_t$. Now for a metastable polaron state with a long lifetime, we
consider time-dependent amplitudes of the form
$\alpha_{0}^{(\vect{p})}(t) = \alpha_{0}^{(\vect{p})}(0) e^{-iEt
  -\Gamma t}$ and $\alpha_{\vect{k}\vect{q}}^{(\vect{p})}(t) =
\alpha_{\vect{k}\vect{q}}^{(\vect{p})}(0) e^{-iEt +\Gamma t}$, where
the decay rate $\Gamma \ll |E|$. Note that we require
$\alpha_{\vect{k}\vect{q}}^{(\vect{p})}$ to grow exponentially with
time while $\alpha_{0}^{(\vect{p})}$ decreases exponentially so that
the normalization condition $|\alpha_0^{(\vect{p})}(t)|^2 +
\sum_{\vect{k}\vect{q}}
|\alpha_{\vect{k}\vect{q}}^{(\vect{p})}(t)|^2=1$ is preserved up to
leading order in $\Gamma/E$. Inserting these amplitudes into the
dynamical equations for the polaron then yields:
\begin{align}\notag
E - i\Gamma -\epsilon_{\p\down}& =\sum_\q\left[\frac1
g+\sum_\k \frac{1}{E_{\p\q\k}-i\Gamma}\right]^{-1} \\ \notag
& \equiv \Sigma(\vect{p},E+i\Gamma)
\end{align}
which corresponds exactly to the condition for the quasiparticle pole
within the diagrammatic approach~\cite{Massignan:2011fk}.  Thus, we
obtain the usual equations for the energy and the decay rate of the
repulsive polaron:
\begin{align} \notag
E & = \epsilon_{\p\down}+\Re[\Sigma(\vect{p},E+i0)] \\ \notag
\Gamma & \sim - \Im[\Sigma(\vect{p},E+i0)]
%-Z \Im[\Sigma(\vect{p},E+i0)]
\end{align}
In principle, this approach can also be used to study the metastable
states proximate to the binding transitions~\cite{bruun2010}.  Note
that energy and momentum conservation restricts the possible decay
channels and means that a variational description which includes the
metastability needs necessarily to include dressing by extra
particle-hole pairs.

\section{\label{sec:conclusion} {Concluding remarks}}
This work presents a thorough investigation of the highly polarized
limit of a 2D Fermi gas, which can be modelled as a $\down$ impurity
in a $\up$ Fermi sea.  We have analyzed the possible states (polaron,
molecule, trimer...) that the impurity can form and we have
constructed the single-impurity phase diagram using two different
levels of approximation for the impurity wave function.  The simple
``undressed'' wave functions in Sec.~\ref{sec:bare} give us insight
into the few-body correlations of the system and allow us to make
contact with standard mean-field approaches for the spin-imbalanced
Fermi gas. The wave functions dressed with one particle-hole
excitation in Sec.~\ref{sec:dressed} contain the correct first-order
correction to the impurity energy due to interactions with the medium,
and they yield a phase diagram that should be qualitatively, if not
quantitatively, accurate across the full range of interactions
$\eta$. Indeed, our study of the polaron state with two particle-hole
pairs in Sec.~\ref{sec:2ph} suggests that our variational approach is
reasonable even in the regime of strong interactions $\eta \sim 0$.
To better model experiment, the variational calculation may also be
extended to properly include the transverse harmonic confinement ---
for further details, we refer the reader to Ref.~\cite{Levinsen2012}.

Our results show that the trimer is strongly favored by the Fermi sea
since the FFLO molecule is unstable to binding an extra $\up$ fermion,
like in 3D.  However, there is still the possibility of observing the
FFLO phase at lower mass ratios $1.8 \lesssim r \lesssim 2.5$. A
remaining question is what happens to the trimer phase when there is a
finite density of impurities. Naively, one might expect a Fermi liquid
of trimers if they are sufficiently tightly bound, but the fact that
they have finite angular momentum could impact the properties of this
liquid phase. There is also the question of whether the
single-impurity binding transitions are thermodynamically stable ---
we have already shown that the polaron-molecule transition for equal
masses is preempted by a SF-N transition at zero temperature. Such
single-impurity transitions may nonetheless survive at finite
temperature, in which case a finite density of molecules would
correspond to a normal phase of preformed pairs.

Finally, we have computed rigorous upper and lower bounds for the
position of the polaron-molecule transition in the case of equal
masses. This should provide a useful benchmark for future experimental
and theoretical work on this topic.

%%%%%%%%%%%%%%%%%%%%%%%%%%%%%%%%%%%%%%%%%%%%%%%%%
\acknowledgments We gratefully acknowledge fruitful discussion with
Stefan Baur, Nigel Cooper, David Huse, Michael K\"ohl, Francesca
Marchetti, Pietro Massignan, Charles Mathy, and Vudtiwat
Ngampruetikorn. We also thank Michael K\"ohl for sharing his
experimental data. MMP acknowledges support from the EPSRC under Grant
No.\ EP/H00369X/2.  JL acknowledges support from a Carlsberg
Foundation Fellowship and a Marie Curie Intra European grant within
the 7th European Community Framework Programme.

%%%%%%%%%%%%%%%%%%%%%%%%%%%%%%%%%%%%%%%%%%%%%%%%%

\appendix

\section{Bound states in the $N+1$ problem \label{app:a}}
In this Appendix we demonstrate how the equations for the binding
energy of a composite consisting of $N$ majority particles and 1
minority particle may be obtained by a diagrammatic technique. This is
an alternative to the variational method described in the main text.

Consider first the undressed wavefunctions, the subject of Section
\ref{sec:bare}. For simplicity of notation, we will here consider the
quasiparticle state at rest. The generalization to finite momentum may
be easily achieved. The $\up$ particles have momenta
$\k_1,\ldots,\k_N$ and corresponding energies $\ekone,\ldots,\ekN$,
while the impurity has momentum $-\sum_i\k_i$ and energy $E-\sum_i
\eki$, such that the total energy is simply $E$.

The sum of diagrams with $N+1$ incoming particles where the impurity
interacts first with the $\up$ particle of momentum $\k_1$ will be
denoted $f_{\k_2\ldots\k_N}$. The function $f$ does not depend on $\k_1$
since the initial interaction depends solely on the total momentum of
the two particles. As the majority particles are fermions, $f$ is
anti-symmetric in its indices. The specific ordering of indices is of
course arbitrary, however once an ordering is chosen it must be kept
throughout the calculation as this corresponds to a choice of ordering
of operators in Wick's theorem.

The occurence of an $N+1$ particle bound state with binding energy $E$
corresponds to a singularity of $f$ at this energy. For the polaron,
$f$ is simply the bare impurity propagator and the energy is that of
the non-interacting impurity. For the molecule, $f$ is the pair
propagator in the medium, denoted $T_2$. For $N\geq2$ the singularity
appears from the summation of an infinite number of diagrams, and may
be found by solving the integral equation illustrated in
Fig.~\ref{fig:multi0}: The initial interaction between the impurity
and the particle with momentum $\k_1$ is described by a pair
propagator. Subsequently, the impurity interacts with another of the
initial particles. Thus the right hand side contains $N-1$ terms and
the function $f$ satisfies the integral equation
\begin{widetext}
\begin{equation}
f_{\k_2\ldots\k_N}\left[\frac1g+%\frac1\Omega
\sum_{\k_1}\frac1{E_{\k_1\ldots\k_N}-i0}\right]=%\frac1\Omega
\sum_{\k_1}
\frac{f_{\k_1\k_3\ldots\k_N}+f_{\k_2\k_1\k_4\ldots\k_N}+\ldots+f_{\k_2\ldots\k_{N-1}\k_1}}{E_{\k_1\ldots\k_N}-i0}.
\end{equation}
\end{widetext}
where
$E_{\k_1\ldots\k_N}=-E+\epsilon_{\k_1+\ldots+\k_N\down}+\sum_i\eki$
and the factor $-i0$ acts to slightly shift the energy pole into the
lower half of the complex plane. The quantity in brackets on the left
hand side is the inverse pair propagator.

\section{The dressed $N+1$ problem \label{app:b}}

We turn now to the states dressed by one particle-hole excitation as
investigated in Section \ref{sec:dressed}.  Again, we construct first
the sum of all diagrams with $N$ incoming $\up$ particles in which the
impurity interacts first with the $\up$ particle of momentum
$\k_1$. The kinematics is chosen as above. Now we allow the vertex to
be dressed by one particle hole pair and this new vertex is denoted
$J_{\k_2\ldots\k_N}$. Again, the initial interaction is described
through a pair propagator. Then, in addition to terms of the same form
as above where the impurity interacts next with another of the initial
particles, it may also interact next with a particle from the Fermi
sea. This is illustrated in Fig.~\ref{fig:multi1}a where the sign
arises from the fermion loop. The vertex $G_{\q\k_1\ldots\k_N}$ is the
sum of all diagrams with one incoming $\down$ particle and $N+1$
incoming $\up$ particles, the initial interaction being between the
impurity and the particle having momentum $\q$ and energy $\eq$. The
equation in Fig.~\ref{fig:multi1}a is
\begin{widetext}
\begin{equation}
J_{\k_2\ldots\k_N}\left[\frac1g+%\frac1\Omega
\sum_{\k_1}\frac1{E_{\k_1\ldots\k_N}-i0}\right]
=%\frac1\Omega
\sum_{\k_1}
\frac{J_{\k_1\k_3\ldots\k_N}+J_{\k_2\k_1\k_4\ldots\k_N}+\ldots+J_{\k_2\ldots\k_{N-1}\k_1}}{E_{\k_1\ldots\k_N}-i0}
-%\frac1\Omega
\sum_{\k_1\q}\frac{G_{\q\k_1\ldots\k_n}}{E_{\k_1\ldots\k_n}-i0}.
\label{eq:Japp}
\end{equation}
\begin{figure}
\begin{center}
\includegraphics[width=.7\linewidth,angle=0]{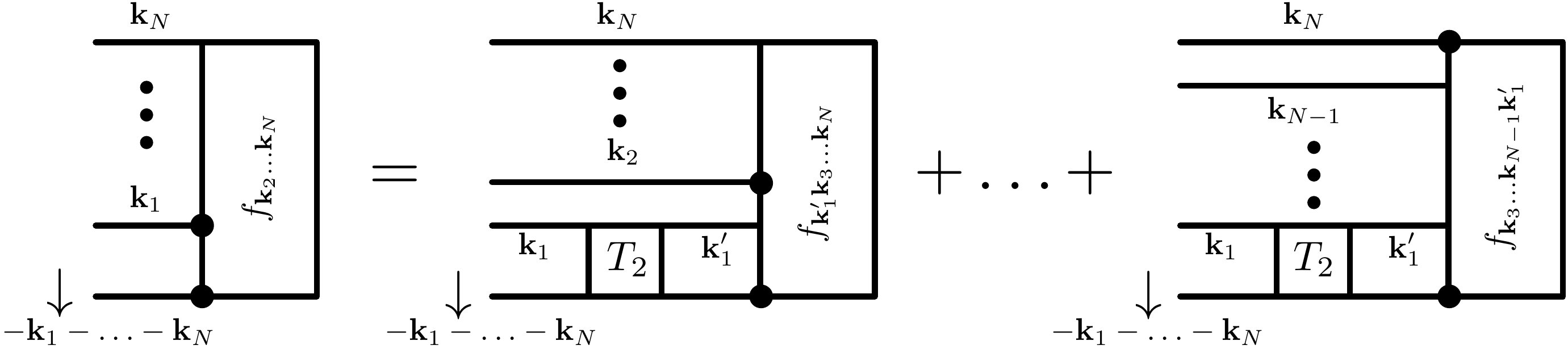}
\end{center}
\caption{The diagrams which lead to the binding energy of the $N+1$
  composite. Black dots on vertices indicate that the corresponding
  particles interact first inside the vertex $f$.}
\label{fig:multi0}
\end{figure}
\begin{figure}
\begin{center}
\includegraphics[width=1\linewidth,angle=0]{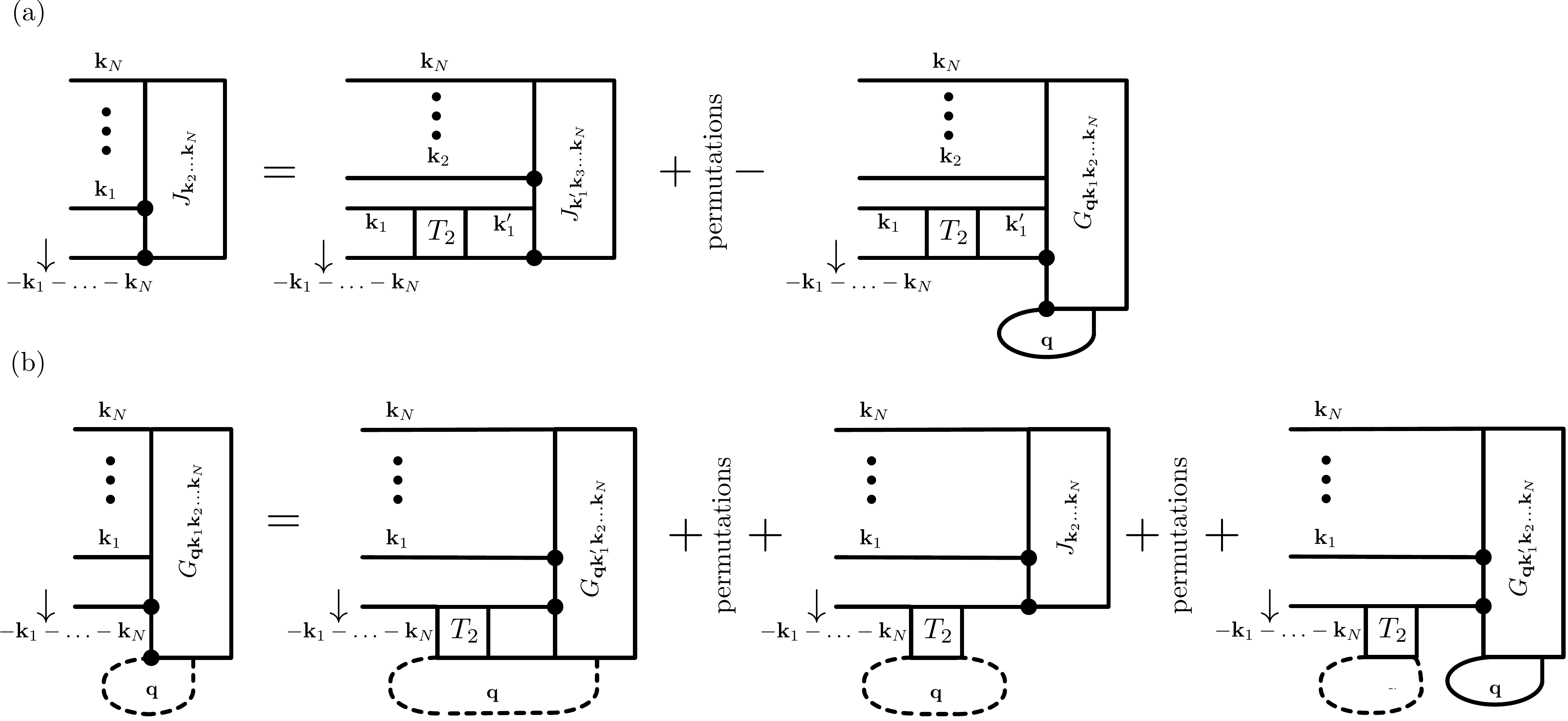}
\end{center}
\caption{The equations which give the binding energy of the $N+1$
  composite including dressing by one particle-hole pair. Black dots
  on vertices indicate that the corresponding particles interact first
  inside the vertex. (a) The equation for the vertex $J$, there are
  $N-1$ diagrams of the first type on the r.h.s. corresponding to
  interactions between the impurity and one of the particles with
  momentum $\k_2,\ldots,\k_N$ and 1 of the second type. (b) The
  equation for $G$. The first two types of diagrams on the
  r.h.s. appear $N$ times corresponding to interactions between the
  impurity and one of the particles with momentum
  $\k_1,\ldots,\k_N$. Dashed lines indicate that these are the loops
  which are closed upon insertion of $G$ in the diagrams in (a).}
\label{fig:multi1}
\end{figure}
\end{widetext}
The vertex $G$ is constructed as follows: The initial interaction
between the impurity and the particle with momentum $\q$ is again
described by a pair propagator. After this interaction, the $\up$
particle which partakes in this repeated interaction is either
disconnected or connected from the remaining particles. If it is
connected then the impurity interacts next with one of the particles
with momentum $\k_1,\ldots,\k_N$, as the vertex is only dressed by one
particle hole pair. This interaction is again described by the vertex
$G$. If disconnected, the impurity can interact first with one of the
particles with momentum $\k_1,\ldots,\k_N$ and this interaction is
given by the vertex $J$, or it can interact first with a particle from
the Fermi sea. This results in the equation for $G$:
\begin{widetext}
\begin{eqnarray}
G_{\q\k_1\ldots\k_N}
\left[\frac1g+%\frac1\Omega
\sum_{\k}\frac1{E_{\q\k_1\ldots\k_N\k}-i0}\right]
& = & 
%\frac1\Omega
\sum_{\k}
\frac{G_{\q\k\k_2\ldots\k_N}+G_{\q\k_1\k\k_3\ldots\k_N}+\ldots+
G_{\q\k_1\ldots\k_{N-1}\k}}{E_{\q\k_1\ldots\k_N\k}-i0}
\nn \\ && \hspace{-25mm}-\frac{J_{\k_2\ldots\k_N}
-J_{\k_1\k_3\ldots\k_N}+J_{\k_1\k_2\k_4\ldots\k_N}-\ldots+(-1)^{N-1}J_{\k_1\ldots\k_{N-1}}}
{E_{\k_1\ldots\k_N}-i0}
-\frac{%\frac1\Omega
\sum_{\q'} G_{\q'\k_1\ldots\k_N}}{E_{\k_1\ldots\k_N}-i0},
\label{eq:Gapp}
\end{eqnarray}
\end{widetext}
where
$E_{\q\k_1\ldots\k_N\k}=-E+\epsilon_{\q-\k_1\ldots-\k_N-\k\down}
+\sum_i\eki+\ek-\eq$. Eq.~(\ref{eq:Gapp}) is illustrated
in Fig.~\ref{fig:multi1}b where the dotted lines, indicating how the
loops are closed, should only be taken as a guide to the eye, as the
integration over momentum in these loops is not performed until the
insertion of this equation in Eq.~(\ref{eq:Japp}).

\bibliography{Ref2D,2DRefs,morerefs}

\begin{thebibliography}{57}
\expandafter\ifx\csname natexlab\endcsname\relax\def\natexlab#1{#1}\fi
\expandafter\ifx\csname bibnamefont\endcsname\relax
  \def\bibnamefont#1{#1}\fi
\expandafter\ifx\csname bibfnamefont\endcsname\relax
  \def\bibfnamefont#1{#1}\fi
\expandafter\ifx\csname citenamefont\endcsname\relax
  \def\citenamefont#1{#1}\fi
\expandafter\ifx\csname url\endcsname\relax
  \def\url#1{\texttt{#1}}\fi
\expandafter\ifx\csname urlprefix\endcsname\relax\def\urlprefix{URL }\fi
\providecommand{\bibinfo}[2]{#2}
\providecommand{\eprint}[2][]{\url{#2}}

\bibitem[{\citenamefont{Martiyanov et~al.}(2010)\citenamefont{Martiyanov,
  Makhalov, and Turlapov}}]{2DFermi_expt}
\bibinfo{author}{\bibfnamefont{K.}~\bibnamefont{Martiyanov}},
  \bibinfo{author}{\bibfnamefont{V.}~\bibnamefont{Makhalov}}, \bibnamefont{and}
  \bibinfo{author}{\bibfnamefont{A.}~\bibnamefont{Turlapov}},
  \bibinfo{journal}{Phys. Rev. Lett.} \textbf{\bibinfo{volume}{105}},
  \bibinfo{pages}{030404} (\bibinfo{year}{2010}).

\bibitem[{\citenamefont{Fr\"ohlich et~al.}(2011)\citenamefont{Fr\"ohlich, Feld,
  Vogt, Koschorreck, Zwerger, and K\"ohl}}]{PhysRevLett.106.105301}
\bibinfo{author}{\bibfnamefont{B.}~\bibnamefont{Fr\"ohlich}},
  \bibinfo{author}{\bibfnamefont{M.}~\bibnamefont{Feld}},
  \bibinfo{author}{\bibfnamefont{E.}~\bibnamefont{Vogt}},
  \bibinfo{author}{\bibfnamefont{M.}~\bibnamefont{Koschorreck}},
  \bibinfo{author}{\bibfnamefont{W.}~\bibnamefont{Zwerger}}, \bibnamefont{and}
  \bibinfo{author}{\bibfnamefont{M.}~\bibnamefont{K\"ohl}},
  \bibinfo{journal}{Phys. Rev. Lett.} \textbf{\bibinfo{volume}{106}},
  \bibinfo{pages}{105301} (\bibinfo{year}{2011}).

\bibitem[{\citenamefont{Dyke et~al.}(2011)\citenamefont{Dyke, Kuhnle, Whitlock,
  Hu, Mark, Hoinka, Lingham, Hannaford, and Vale}}]{Dyke2011}
\bibinfo{author}{\bibfnamefont{P.}~\bibnamefont{Dyke}},
  \bibinfo{author}{\bibfnamefont{E.~D.} \bibnamefont{Kuhnle}},
  \bibinfo{author}{\bibfnamefont{S.}~\bibnamefont{Whitlock}},
  \bibinfo{author}{\bibfnamefont{H.}~\bibnamefont{Hu}},
  \bibinfo{author}{\bibfnamefont{M.}~\bibnamefont{Mark}},
  \bibinfo{author}{\bibfnamefont{S.}~\bibnamefont{Hoinka}},
  \bibinfo{author}{\bibfnamefont{M.}~\bibnamefont{Lingham}},
  \bibinfo{author}{\bibfnamefont{P.}~\bibnamefont{Hannaford}},
  \bibnamefont{and} \bibinfo{author}{\bibfnamefont{C.~J.} \bibnamefont{Vale}},
  \bibinfo{journal}{Phys. Rev. Lett.} \textbf{\bibinfo{volume}{106}},
  \bibinfo{pages}{105304} (\bibinfo{year}{2011}).

\bibitem[{\citenamefont{{Feld} et~al.}(2011)\citenamefont{{Feld},
  {Fr{\"o}hlich}, {Vogt}, {Koschorreck}, and {K{\"o}hl}}}]{2011Natur.480...75F}
\bibinfo{author}{\bibfnamefont{M.}~\bibnamefont{{Feld}}},
  \bibinfo{author}{\bibfnamefont{B.}~\bibnamefont{{Fr{\"o}hlich}}},
  \bibinfo{author}{\bibfnamefont{E.}~\bibnamefont{{Vogt}}},
  \bibinfo{author}{\bibfnamefont{M.}~\bibnamefont{{Koschorreck}}},
  \bibnamefont{and}
  \bibinfo{author}{\bibfnamefont{M.}~\bibnamefont{{K{\"o}hl}}},
  \bibinfo{journal}{\nat} \textbf{\bibinfo{volume}{480}}, \bibinfo{pages}{75}
  (\bibinfo{year}{2011}).

\bibitem[{\citenamefont{Sommer et~al.}(2012)\citenamefont{Sommer, Cheuk, Ku,
  Bakr, and Zwierlein}}]{sommer2011_2D}
\bibinfo{author}{\bibfnamefont{A.~T.} \bibnamefont{Sommer}},
  \bibinfo{author}{\bibfnamefont{L.~W.} \bibnamefont{Cheuk}},
  \bibinfo{author}{\bibfnamefont{M.~J.~H.} \bibnamefont{Ku}},
  \bibinfo{author}{\bibfnamefont{W.~S.} \bibnamefont{Bakr}}, \bibnamefont{and}
  \bibinfo{author}{\bibfnamefont{M.~W.} \bibnamefont{Zwierlein}},
  \bibinfo{journal}{Phys. Rev. Lett.} \textbf{\bibinfo{volume}{108}},
  \bibinfo{pages}{045302} (\bibinfo{year}{2012}).

\bibitem[{\citenamefont{{Koschorreck} et~al.}(2012)\citenamefont{{Koschorreck},
  {Pertot}, {Vogt}, {Fr{\"o}hlich}, {Feld}, and {K{\"o}hl}}}]{Koschorreck2012}
\bibinfo{author}{\bibfnamefont{M.}~\bibnamefont{{Koschorreck}}},
  \bibinfo{author}{\bibfnamefont{D.}~\bibnamefont{{Pertot}}},
  \bibinfo{author}{\bibfnamefont{E.}~\bibnamefont{{Vogt}}},
  \bibinfo{author}{\bibfnamefont{B.}~\bibnamefont{{Fr{\"o}hlich}}},
  \bibinfo{author}{\bibfnamefont{M.}~\bibnamefont{{Feld}}}, \bibnamefont{and}
  \bibinfo{author}{\bibfnamefont{M.}~\bibnamefont{{K{\"o}hl}}},
  \bibinfo{journal}{\nat} \textbf{\bibinfo{volume}{485}}, \bibinfo{pages}{619}
  (\bibinfo{year}{2012}).

\bibitem[{\citenamefont{Zhang et~al.}(2012)\citenamefont{Zhang, Ong, Arakelyan,
  and Thomas}}]{Zhang:2012uq}
\bibinfo{author}{\bibfnamefont{Y.}~\bibnamefont{Zhang}},
  \bibinfo{author}{\bibfnamefont{W.}~\bibnamefont{Ong}},
  \bibinfo{author}{\bibfnamefont{I.}~\bibnamefont{Arakelyan}},
  \bibnamefont{and} \bibinfo{author}{\bibfnamefont{J.~E.}
  \bibnamefont{Thomas}}, \bibinfo{journal}{Phys. Rev. Lett.}
  \textbf{\bibinfo{volume}{108}}, \bibinfo{pages}{235302}
  (\bibinfo{year}{2012}).

\bibitem[{\citenamefont{Croxall et~al.}(2008)\citenamefont{Croxall, Das~Gupta,
  Nicoll, Thangaraj, Beere, Farrer, Ritchie, and Pepper}}]{Croxall2008}
\bibinfo{author}{\bibfnamefont{A.~F.} \bibnamefont{Croxall}},
  \bibinfo{author}{\bibfnamefont{K.}~\bibnamefont{Das~Gupta}},
  \bibinfo{author}{\bibfnamefont{C.~A.} \bibnamefont{Nicoll}},
  \bibinfo{author}{\bibfnamefont{M.}~\bibnamefont{Thangaraj}},
  \bibinfo{author}{\bibfnamefont{H.~E.} \bibnamefont{Beere}},
  \bibinfo{author}{\bibfnamefont{I.}~\bibnamefont{Farrer}},
  \bibinfo{author}{\bibfnamefont{D.~A.} \bibnamefont{Ritchie}},
  \bibnamefont{and} \bibinfo{author}{\bibfnamefont{M.}~\bibnamefont{Pepper}},
  \bibinfo{journal}{Phys. Rev. Lett.} \textbf{\bibinfo{volume}{101}},
  \bibinfo{pages}{246801} (\bibinfo{year}{2008}).

\bibitem[{\citenamefont{Seamons et~al.}(2009)\citenamefont{Seamons, Morath,
  Reno, and Lilly}}]{Seamons2009}
\bibinfo{author}{\bibfnamefont{J.~A.} \bibnamefont{Seamons}},
  \bibinfo{author}{\bibfnamefont{C.~P.} \bibnamefont{Morath}},
  \bibinfo{author}{\bibfnamefont{J.~L.} \bibnamefont{Reno}}, \bibnamefont{and}
  \bibinfo{author}{\bibfnamefont{M.~P.} \bibnamefont{Lilly}},
  \bibinfo{journal}{Phys. Rev. Lett.} \textbf{\bibinfo{volume}{102}},
  \bibinfo{pages}{026804} (\bibinfo{year}{2009}).

\bibitem[{\citenamefont{Norman}(2011)}]{Norman2011}
\bibinfo{author}{\bibfnamefont{M.~R.} \bibnamefont{Norman}},
  \bibinfo{journal}{Science} \textbf{\bibinfo{volume}{332}},
  \bibinfo{pages}{196} (\bibinfo{year}{2011}).

\bibitem[{\citenamefont{Fr{\"o}hlich}(1954)}]{frohlich1954}
\bibinfo{author}{\bibfnamefont{H.}~\bibnamefont{Fr{\"o}hlich}},
  \bibinfo{journal}{Advances in Physics} \textbf{\bibinfo{volume}{3}},
  \bibinfo{pages}{325} (\bibinfo{year}{1954}).

\bibitem[{\citenamefont{Prokof'ev and
  Svistunov}(2008{\natexlab{a}})}]{prokofiev2008}
\bibinfo{author}{\bibfnamefont{N.}~\bibnamefont{Prokof'ev}} \bibnamefont{and}
  \bibinfo{author}{\bibfnamefont{B.}~\bibnamefont{Svistunov}},
  \bibinfo{journal}{Phys. Rev. B} \textbf{\bibinfo{volume}{77}},
  \bibinfo{eid}{020408} (\bibinfo{year}{2008}{\natexlab{a}}).

\bibitem[{\citenamefont{Prokof'ev and
  Svistunov}(2008{\natexlab{b}})}]{prokofiev2008_2}
\bibinfo{author}{\bibfnamefont{N.~V.} \bibnamefont{Prokof'ev}}
  \bibnamefont{and} \bibinfo{author}{\bibfnamefont{B.~V.}
  \bibnamefont{Svistunov}}, \bibinfo{journal}{Phys. Rev. B}
  \textbf{\bibinfo{volume}{77}}, \bibinfo{eid}{125101}
  (\bibinfo{year}{2008}{\natexlab{b}}).

\bibitem[{\citenamefont{Parish et~al.}(2007{\natexlab{a}})\citenamefont{Parish,
  Marchetti, Lamacraft, and Simons}}]{parish2007}
\bibinfo{author}{\bibfnamefont{M.~M.} \bibnamefont{Parish}},
  \bibinfo{author}{\bibfnamefont{F.~M.} \bibnamefont{Marchetti}},
  \bibinfo{author}{\bibfnamefont{A.}~\bibnamefont{Lamacraft}},
  \bibnamefont{and} \bibinfo{author}{\bibfnamefont{B.~D.}
  \bibnamefont{Simons}}, \bibinfo{journal}{Nature Phys.}
  \textbf{\bibinfo{volume}{3}}, \bibinfo{pages}{124}
  (\bibinfo{year}{2007}{\natexlab{a}}).

\bibitem[{\citenamefont{Sheehy and Radzihovsky}(2007)}]{sheehy2007}
\bibinfo{author}{\bibfnamefont{D.~E.} \bibnamefont{Sheehy}} \bibnamefont{and}
  \bibinfo{author}{\bibfnamefont{L.}~\bibnamefont{Radzihovsky}},
  \bibinfo{journal}{Annals of Physics} \textbf{\bibinfo{volume}{322}},
  \bibinfo{pages}{1790} (\bibinfo{year}{2007}).

\bibitem[{\citenamefont{Chevy}(2006)}]{chevy2006_2}
\bibinfo{author}{\bibfnamefont{F.}~\bibnamefont{Chevy}},
  \bibinfo{journal}{Phys. Rev. A} \textbf{\bibinfo{volume}{74}},
  \bibinfo{pages}{063628} (\bibinfo{year}{2006}).

\bibitem[{\citenamefont{Combescot et~al.}(2009)\citenamefont{Combescot, Giraud,
  and Leyronas}}]{combescot2009}
\bibinfo{author}{\bibfnamefont{R.}~\bibnamefont{Combescot}},
  \bibinfo{author}{\bibfnamefont{S.}~\bibnamefont{Giraud}}, \bibnamefont{and}
  \bibinfo{author}{\bibfnamefont{X.}~\bibnamefont{Leyronas}},
  \bibinfo{journal}{Europhys. Lett.} \textbf{\bibinfo{volume}{88}},
  \bibinfo{pages}{60007} (\bibinfo{year}{2009}).

\bibitem[{\citenamefont{Mora and Chevy}(2009)}]{mora2009}
\bibinfo{author}{\bibfnamefont{C.}~\bibnamefont{Mora}} \bibnamefont{and}
  \bibinfo{author}{\bibfnamefont{F.}~\bibnamefont{Chevy}},
  \bibinfo{journal}{Phys. Rev. A} \textbf{\bibinfo{volume}{80}},
  \bibinfo{eid}{033607} (\bibinfo{year}{2009}).

\bibitem[{\citenamefont{Punk et~al.}(2009)\citenamefont{Punk, Dumitrescu, and
  Zwerger}}]{punk2009}
\bibinfo{author}{\bibfnamefont{M.}~\bibnamefont{Punk}},
  \bibinfo{author}{\bibfnamefont{P.~T.} \bibnamefont{Dumitrescu}},
  \bibnamefont{and} \bibinfo{author}{\bibfnamefont{W.}~\bibnamefont{Zwerger}},
  \bibinfo{journal}{Phys. Rev. A} \textbf{\bibinfo{volume}{80}},
  \bibinfo{eid}{053605} (\bibinfo{year}{2009}).

\bibitem[{\citenamefont{Mathy et~al.}(2011)\citenamefont{Mathy, Parish, and
  Huse}}]{Mathy:2011ys}
\bibinfo{author}{\bibfnamefont{C.~J.~M.} \bibnamefont{Mathy}},
  \bibinfo{author}{\bibfnamefont{M.~M.} \bibnamefont{Parish}},
  \bibnamefont{and} \bibinfo{author}{\bibfnamefont{D.~A.} \bibnamefont{Huse}},
  \bibinfo{journal}{Phys. Rev. Lett} \textbf{\bibinfo{volume}{106}},
  \bibinfo{pages}{166404} (\bibinfo{year}{2011}).

\bibitem[{\citenamefont{Z\"{o}llner et~al.}(2011)\citenamefont{Z\"{o}llner,
  Bruun, and Pethick}}]{Zollner:2011fk}
\bibinfo{author}{\bibfnamefont{S.}~\bibnamefont{Z\"{o}llner}},
  \bibinfo{author}{\bibfnamefont{G.~M.} \bibnamefont{Bruun}}, \bibnamefont{and}
  \bibinfo{author}{\bibfnamefont{C.~J.} \bibnamefont{Pethick}},
  \bibinfo{journal}{Phys. Rev. A} \textbf{\bibinfo{volume}{83}},
  \bibinfo{pages}{021603(R)} (\bibinfo{year}{2011}).

\bibitem[{\citenamefont{Klawunn and Recati}(2011)}]{Klawunn:2011fk}
\bibinfo{author}{\bibfnamefont{M.}~\bibnamefont{Klawunn}} \bibnamefont{and}
  \bibinfo{author}{\bibfnamefont{A.}~\bibnamefont{Recati}},
  \bibinfo{journal}{Phys. Rev. A} \textbf{\bibinfo{volume}{84}},
  \bibinfo{pages}{033607} (\bibinfo{year}{2011}).

\bibitem[{\citenamefont{Parish}(2011)}]{Parish:2011vn}
\bibinfo{author}{\bibfnamefont{M.~M.} \bibnamefont{Parish}},
  \bibinfo{journal}{Phys. Rev. A} \textbf{\bibinfo{volume}{83}},
  \bibinfo{pages}{051603} (\bibinfo{year}{2011}).

\bibitem[{\citenamefont{Levinsen and Baur}(2012)}]{Levinsen2012}
\bibinfo{author}{\bibfnamefont{J.}~\bibnamefont{Levinsen}} \bibnamefont{and}
  \bibinfo{author}{\bibfnamefont{S.~K.} \bibnamefont{Baur}},
  \bibinfo{journal}{Phys. Rev. A} \textbf{\bibinfo{volume}{86}},
  \bibinfo{pages}{041602} (\bibinfo{year}{2012}).

\bibitem[{\citenamefont{Pricoupenko and Pedri}(2010)}]{2Dtrimer}
\bibinfo{author}{\bibfnamefont{L.}~\bibnamefont{Pricoupenko}} \bibnamefont{and}
  \bibinfo{author}{\bibfnamefont{P.}~\bibnamefont{Pedri}},
  \bibinfo{journal}{Phys. Rev. A} \textbf{\bibinfo{volume}{82}},
  \bibinfo{pages}{033625} (\bibinfo{year}{2010}).

\bibitem[{\citenamefont{Levinsen and Parish}(2012)}]{Levinsen2013}
\bibinfo{author}{\bibfnamefont{J.}~\bibnamefont{Levinsen}} \bibnamefont{and}
  \bibinfo{author}{\bibfnamefont{M.~M.} \bibnamefont{Parish}},
  \bibinfo{journal}{arXiv:1207.0459}  (\bibinfo{year}{2012}).

\bibitem[{\citenamefont{Fulde and Ferrell}(1964)}]{PhysRev.135.A550}
\bibinfo{author}{\bibfnamefont{P.}~\bibnamefont{Fulde}} \bibnamefont{and}
  \bibinfo{author}{\bibfnamefont{R.~A.} \bibnamefont{Ferrell}},
  \bibinfo{journal}{Phys. Rev.} \textbf{\bibinfo{volume}{135}},
  \bibinfo{pages}{A550} (\bibinfo{year}{1964}).

\bibitem[{\citenamefont{{Larkin} and {Ovchinnikov}}(1967)}]{1965JETP...20.762L}
\bibinfo{author}{\bibfnamefont{A.~I.} \bibnamefont{{Larkin}}} \bibnamefont{and}
  \bibinfo{author}{\bibfnamefont{Y.~N.} \bibnamefont{{Ovchinnikov}}},
  \bibinfo{journal}{Soviet JETP} \textbf{\bibinfo{volume}{20}},
  \bibinfo{pages}{762} (\bibinfo{year}{1967}).

\bibitem[{\citenamefont{Levinsen et~al.}(2009)\citenamefont{Levinsen, Tiecke,
  Walraven, and Petrov}}]{PhysRevLett.103.153202}
\bibinfo{author}{\bibfnamefont{J.}~\bibnamefont{Levinsen}},
  \bibinfo{author}{\bibfnamefont{T.~G.} \bibnamefont{Tiecke}},
  \bibinfo{author}{\bibfnamefont{J.~T.~M.} \bibnamefont{Walraven}},
  \bibnamefont{and} \bibinfo{author}{\bibfnamefont{D.~S.}
  \bibnamefont{Petrov}}, \bibinfo{journal}{Phys. Rev. Lett.}
  \textbf{\bibinfo{volume}{103}}, \bibinfo{pages}{153202}
  (\bibinfo{year}{2009}).

\bibitem[{\citenamefont{{Ngampruetikorn}
  et~al.}(2012)\citenamefont{{Ngampruetikorn}, {Parish}, and
  {Levinsen}}}]{Ngampruetikorn2013}
\bibinfo{author}{\bibfnamefont{V.}~\bibnamefont{{Ngampruetikorn}}},
  \bibinfo{author}{\bibfnamefont{M.~M.} \bibnamefont{{Parish}}},
  \bibnamefont{and}
  \bibinfo{author}{\bibfnamefont{J.}~\bibnamefont{{Levinsen}}},
  \bibinfo{journal}{arXiv:1211.6805}  (\bibinfo{year}{2012}).

\bibitem[{\citenamefont{Combescot et~al.}(2007)\citenamefont{Combescot, Recati,
  Lobo, and Chevy}}]{Combescot:2007bh}
\bibinfo{author}{\bibfnamefont{R.}~\bibnamefont{Combescot}},
  \bibinfo{author}{\bibfnamefont{A.}~\bibnamefont{Recati}},
  \bibinfo{author}{\bibfnamefont{C.}~\bibnamefont{Lobo}}, \bibnamefont{and}
  \bibinfo{author}{\bibfnamefont{F.}~\bibnamefont{Chevy}},
  \bibinfo{journal}{Phys. Rev. Lett.} \textbf{\bibinfo{volume}{98}},
  \bibinfo{pages}{180402} (\bibinfo{year}{2007}).

\bibitem[{\citenamefont{Bertaina and Giorgini}(2011)}]{PhysRevLett.106.110403}
\bibinfo{author}{\bibfnamefont{G.}~\bibnamefont{Bertaina}} \bibnamefont{and}
  \bibinfo{author}{\bibfnamefont{S.}~\bibnamefont{Giorgini}},
  \bibinfo{journal}{Phys. Rev. Lett.} \textbf{\bibinfo{volume}{106}},
  \bibinfo{pages}{110403} (\bibinfo{year}{2011}).

\bibitem[{\citenamefont{Landau and Lifshitz}(1981)}]{LL}
\bibinfo{author}{\bibfnamefont{L.~D.} \bibnamefont{Landau}} \bibnamefont{and}
  \bibinfo{author}{\bibfnamefont{E.~M.} \bibnamefont{Lifshitz}},
  \emph{\bibinfo{title}{Quantum Mechanics}}
  (\bibinfo{publisher}{Butterworth-Heinemann}, \bibinfo{address}{Oxford, UK},
  \bibinfo{year}{1981}).

\bibitem[{\citenamefont{Bloom}(1975)}]{PhysRevB.12.125}
\bibinfo{author}{\bibfnamefont{P.}~\bibnamefont{Bloom}},
  \bibinfo{journal}{Phys. Rev. B} \textbf{\bibinfo{volume}{12}},
  \bibinfo{pages}{125} (\bibinfo{year}{1975}).

\bibitem[{\citenamefont{Parish et~al.}(2007{\natexlab{b}})\citenamefont{Parish,
  Marchetti, Lamacraft, and Simons}}]{parish2007_2}
\bibinfo{author}{\bibfnamefont{M.~M.} \bibnamefont{Parish}},
  \bibinfo{author}{\bibfnamefont{F.~M.} \bibnamefont{Marchetti}},
  \bibinfo{author}{\bibfnamefont{A.}~\bibnamefont{Lamacraft}},
  \bibnamefont{and} \bibinfo{author}{\bibfnamefont{B.~D.}
  \bibnamefont{Simons}}, \bibinfo{journal}{Phys. Rev. Lett.}
  \textbf{\bibinfo{volume}{98}}, \bibinfo{pages}{160402}
  (\bibinfo{year}{2007}{\natexlab{b}}).

\bibitem[{\citenamefont{Delves and Mohamed}(1985)}]{numerics-book}
\bibinfo{author}{\bibfnamefont{L.~M.} \bibnamefont{Delves}} \bibnamefont{and}
  \bibinfo{author}{\bibfnamefont{J.~L.} \bibnamefont{Mohamed}},
  \emph{\bibinfo{title}{Computational methods for integral equations}}
  (\bibinfo{publisher}{Cambridge University Press}, \bibinfo{year}{1985}).

\bibitem[{\citenamefont{Castin et~al.}(2010)\citenamefont{Castin, Mora, and
  Pricoupenko}}]{PhysRevLett.105.223201}
\bibinfo{author}{\bibfnamefont{Y.}~\bibnamefont{Castin}},
  \bibinfo{author}{\bibfnamefont{C.}~\bibnamefont{Mora}}, \bibnamefont{and}
  \bibinfo{author}{\bibfnamefont{L.}~\bibnamefont{Pricoupenko}},
  \bibinfo{journal}{Phys. Rev. Lett.} \textbf{\bibinfo{volume}{105}},
  \bibinfo{pages}{223201} (\bibinfo{year}{2010}).

\bibitem[{\citenamefont{Minlos}(1995)}]{Minlos}
\bibinfo{author}{\bibfnamefont{R.}~\bibnamefont{Minlos}}, in
  \emph{\bibinfo{booktitle}{Proceedings of the Workshop on Singular
  Schr\"odinger Operators, Trieste, 29 September-1 October 1994}}, edited by
  \bibinfo{editor}{\bibfnamefont{G.}~\bibnamefont{Dell'Antonio}},
  \bibinfo{editor}{\bibfnamefont{R.}~\bibnamefont{Figari}}, \bibnamefont{and}
  \bibinfo{editor}{\bibfnamefont{A.}~\bibnamefont{Teta}}
  (\bibinfo{publisher}{SISSA}, \bibinfo{address}{Trieste},
  \bibinfo{year}{1995}).

\bibitem[{not({\natexlab{a}})}]{notepower}
\bibinfo{note}{Note, however, that the factor $\pi$ in Eq.~(10) of
  Ref.~\cite{Minlos} should be squared}.

\bibitem[{\citenamefont{Pricoupenko}(2011)}]{Pricoupenko2011}
\bibinfo{author}{\bibfnamefont{L.}~\bibnamefont{Pricoupenko}},
  \bibinfo{journal}{Phys. Rev. A} \textbf{\bibinfo{volume}{83}},
  \bibinfo{pages}{062711} (\bibinfo{year}{2011}).

\bibitem[{\citenamefont{Combescot and Giraud}(2008)}]{combescot2008}
\bibinfo{author}{\bibfnamefont{R.}~\bibnamefont{Combescot}} \bibnamefont{and}
  \bibinfo{author}{\bibfnamefont{S.}~\bibnamefont{Giraud}},
  \bibinfo{journal}{Phys. Rev. Lett.} \textbf{\bibinfo{volume}{101}},
  \bibinfo{eid}{050404} (\bibinfo{year}{2008}).

\bibitem[{\citenamefont{Schmidt et~al.}(2012)\citenamefont{Schmidt, Enss,
  Pietil\"a, and Demler}}]{Schmidt2011}
\bibinfo{author}{\bibfnamefont{R.}~\bibnamefont{Schmidt}},
  \bibinfo{author}{\bibfnamefont{T.}~\bibnamefont{Enss}},
  \bibinfo{author}{\bibfnamefont{V.}~\bibnamefont{Pietil\"a}},
  \bibnamefont{and} \bibinfo{author}{\bibfnamefont{E.}~\bibnamefont{Demler}},
  \bibinfo{journal}{Phys. Rev. A} \textbf{\bibinfo{volume}{85}},
  \bibinfo{pages}{021602} (\bibinfo{year}{2012}).

\bibitem[{\citenamefont{Ngampruetikorn
  et~al.}(2012)\citenamefont{Ngampruetikorn, Levinsen, and
  Parish}}]{Ngampruetikorn2012}
\bibinfo{author}{\bibfnamefont{V.}~\bibnamefont{Ngampruetikorn}},
  \bibinfo{author}{\bibfnamefont{J.}~\bibnamefont{Levinsen}}, \bibnamefont{and}
  \bibinfo{author}{\bibfnamefont{M.~M.} \bibnamefont{Parish}},
  \bibinfo{journal}{Europhys. Lett.} \textbf{\bibinfo{volume}{98}},
  \bibinfo{pages}{30005} (\bibinfo{year}{2012}).

\bibitem[{not({\natexlab{b}})}]{noterescale}
\bibinfo{note}{In the experimental work of Ref.~\cite{Koschorreck2012}, the
  definition $\eb=1/m\ad^2$ is used to relate the binding energy to $\ad$.
  However, in a quasi-2D geometry it is necessary to take into account the,
  approximately harmonic, confining potential $V(z)=\frac12m\omega_z^2z^2$ in
  the $z$-direction in order to obtain the 2D scattering length characterising
  the low-energy scattering: $\ad=\sqrt{\pi/(m B
  \omega_z)}\exp\left[-\sqrt{\pi/2}{\cal F}(\eb/\omega_z)\right]$, with
  $B\approx0.905$, and ${\cal F}(x)=\int_0^\infty\frac{du}{\sqrt{4\pi
  u^3}}\left(1-\frac{e^{-xu}}{\sqrt{[1-\exp(-2u)]/2u}}\right)$~\cite{Petrov:2001fk,RevModPhys.80.885}.
  As argued in Ref.~\cite{Levinsen2012}, this latter definition of $\ad$ yields
  a better agreement between the quasi-2D experiments and the strict 2D theory
  employed in the present work.}

\bibitem[{\citenamefont{Schirotzek et~al.}(2009)\citenamefont{Schirotzek, Wu,
  Sommer, and Zwierlein}}]{PhysRevLett.102.230402}
\bibinfo{author}{\bibfnamefont{A.}~\bibnamefont{Schirotzek}},
  \bibinfo{author}{\bibfnamefont{C.-H.} \bibnamefont{Wu}},
  \bibinfo{author}{\bibfnamefont{A.}~\bibnamefont{Sommer}}, \bibnamefont{and}
  \bibinfo{author}{\bibfnamefont{M.~W.} \bibnamefont{Zwierlein}},
  \bibinfo{journal}{Phys. Rev. Lett.} \textbf{\bibinfo{volume}{102}},
  \bibinfo{pages}{230402} (\bibinfo{year}{2009}).

\bibitem[{\citenamefont{Nascimb\`ene et~al.}(2009)\citenamefont{Nascimb\`ene,
  Navon, Jiang, Tarruell, Teichmann, McKeever, Chevy, and
  Salomon}}]{PhysRevLett.103.170402}
\bibinfo{author}{\bibfnamefont{S.}~\bibnamefont{Nascimb\`ene}},
  \bibinfo{author}{\bibfnamefont{N.}~\bibnamefont{Navon}},
  \bibinfo{author}{\bibfnamefont{K.~J.} \bibnamefont{Jiang}},
  \bibinfo{author}{\bibfnamefont{L.}~\bibnamefont{Tarruell}},
  \bibinfo{author}{\bibfnamefont{M.}~\bibnamefont{Teichmann}},
  \bibinfo{author}{\bibfnamefont{J.}~\bibnamefont{McKeever}},
  \bibinfo{author}{\bibfnamefont{F.}~\bibnamefont{Chevy}}, \bibnamefont{and}
  \bibinfo{author}{\bibfnamefont{C.}~\bibnamefont{Salomon}},
  \bibinfo{journal}{Phys. Rev. Lett.} \textbf{\bibinfo{volume}{103}},
  \bibinfo{pages}{170402} (\bibinfo{year}{2009}).

\bibitem[{\citenamefont{Grimm}(2011)}]{innsbruck}
\bibinfo{author}{\bibfnamefont{R.}~\bibnamefont{Grimm}},
  \bibinfo{howpublished}{talk at BEC 2011, Sant Feliu} (\bibinfo{year}{2011}).

\bibitem[{\citenamefont{Giraud and Combescot}(2009)}]{1dpolaron}
\bibinfo{author}{\bibfnamefont{S.}~\bibnamefont{Giraud}} \bibnamefont{and}
  \bibinfo{author}{\bibfnamefont{R.}~\bibnamefont{Combescot}},
  \bibinfo{journal}{Phys. Rev. A} \textbf{\bibinfo{volume}{79}},
  \bibinfo{pages}{043615} (\bibinfo{year}{2009}).

\bibitem[{Koe()}]{Koehl2012}
\bibinfo{note}{M. K\"ohl, talk at the 2012 APS March meeting.}

\bibitem[{\citenamefont{Bruun and Massignan}(2010)}]{bruun2010}
\bibinfo{author}{\bibfnamefont{G.~M.} \bibnamefont{Bruun}} \bibnamefont{and}
  \bibinfo{author}{\bibfnamefont{P.}~\bibnamefont{Massignan}},
  \bibinfo{journal}{Phys. Rev. Lett.} \textbf{\bibinfo{volume}{105}},
  \bibinfo{pages}{020403} (\bibinfo{year}{2010}).

\bibitem[{\citenamefont{Conduit et~al.}(2008)\citenamefont{Conduit, Conlon, and
  Simons}}]{conduit2008}
\bibinfo{author}{\bibfnamefont{G.~J.} \bibnamefont{Conduit}},
  \bibinfo{author}{\bibfnamefont{P.~H.} \bibnamefont{Conlon}},
  \bibnamefont{and} \bibinfo{author}{\bibfnamefont{B.~D.}
  \bibnamefont{Simons}}, \bibinfo{journal}{Phys. Rev. A}
  \textbf{\bibinfo{volume}{77}}, \bibinfo{pages}{053617}
  (\bibinfo{year}{2008}).

\bibitem[{\citenamefont{Cui and Zhai}(2010)}]{Cui:2010zr}
\bibinfo{author}{\bibfnamefont{X.}~\bibnamefont{Cui}} \bibnamefont{and}
  \bibinfo{author}{\bibfnamefont{H.}~\bibnamefont{Zhai}},
  \bibinfo{journal}{Phys. Rev. A} \textbf{\bibinfo{volume}{81}},
  \bibinfo{pages}{041602} (\bibinfo{year}{2010}).

\bibitem[{\citenamefont{Massignan and Bruun}(2011)}]{Massignan:2011fk}
\bibinfo{author}{\bibfnamefont{P.}~\bibnamefont{Massignan}} \bibnamefont{and}
  \bibinfo{author}{\bibfnamefont{G.~M.} \bibnamefont{Bruun}},
  \bibinfo{journal}{Eur. Phys. J. D}  (\bibinfo{year}{2011}).

\bibitem[{\citenamefont{McLachlan}(1964)}]{McLachlan64}
\bibinfo{author}{\bibfnamefont{A.~D.} \bibnamefont{McLachlan}},
  \bibinfo{journal}{Molecular Physics} \textbf{\bibinfo{volume}{8}},
  \bibinfo{pages}{39} (\bibinfo{year}{1964}).

\bibitem[{\citenamefont{Basile and Elser}(1995)}]{PhysRevE.51.5688}
\bibinfo{author}{\bibfnamefont{A.~G.} \bibnamefont{Basile}} \bibnamefont{and}
  \bibinfo{author}{\bibfnamefont{V.}~\bibnamefont{Elser}},
  \bibinfo{journal}{Phys. Rev. E} \textbf{\bibinfo{volume}{51}},
  \bibinfo{pages}{5688} (\bibinfo{year}{1995}).

\bibitem[{\citenamefont{Petrov and Shlyapnikov}(2001)}]{Petrov:2001fk}
\bibinfo{author}{\bibfnamefont{D.~S.} \bibnamefont{Petrov}} \bibnamefont{and}
  \bibinfo{author}{\bibfnamefont{G.~V.} \bibnamefont{Shlyapnikov}},
  \bibinfo{journal}{Phys. Rev. A} \textbf{\bibinfo{volume}{64}},
  \bibinfo{pages}{012706} (\bibinfo{year}{2001}).

\bibitem[{\citenamefont{Bloch et~al.}(2008)\citenamefont{Bloch, Dalibard, and
  Zwerger}}]{RevModPhys.80.885}
\bibinfo{author}{\bibfnamefont{I.}~\bibnamefont{Bloch}},
  \bibinfo{author}{\bibfnamefont{J.}~\bibnamefont{Dalibard}}, \bibnamefont{and}
  \bibinfo{author}{\bibfnamefont{W.}~\bibnamefont{Zwerger}},
  \bibinfo{journal}{Rev. Mod. Phys.} \textbf{\bibinfo{volume}{80}},
  \bibinfo{pages}{885} (\bibinfo{year}{2008}).

\end{thebibliography}

\end{document}